%% file: main.tex
\pdfoutput=1
\documentclass[conference]{IEEEtran}
\IEEEoverridecommandlockouts

\usepackage{cite}
\usepackage{amsmath,amssymb,amsfonts}
\usepackage{algorithmic}
\usepackage{graphicx}
\usepackage{textcomp}
\usepackage{xcolor}
\usepackage{booktabs}
\usepackage{siunitx}
\usepackage{url}
\usepackage{tikz}
\usetikzlibrary{positioning,arrows.meta,calc,fit}
\def\BibTeX{{\rm B\kern-.05em{\sc i\kern-.025em b}\kern-.08em
    T\kern-.1667em\lower.7ex\hbox{E}\kern-.125emX}}

\graphicspath{{paper_resources/}}

  \newcommand\wsPA{Viralis-Optimalis}\newcommand\wsPB{Artemis}\newcommand\wsPC{Formalis}
  \newcommand\wsRA{Green-Code}\newcommand\wsRB{Cross}\newcommand\wsCu{Customer-Project}

\begin{document}

\title{Re-Reading Is Not Free: Token Traces and Cache-Aware Energy Estimation
for LLM Agents in Software Engineering
}

\author{\IEEEauthorblockN{%
Tina Vartziotis\textsuperscript{1,2,3},
Ketevan Peranidze\textsuperscript{1,4},
Rodopi Kosteli\textsuperscript{5},
Ezgi Sarikayak\textsuperscript{6,7},\\
George Dasoulas\textsuperscript{3},
Michael Keckeisen\textsuperscript{1},
Sotirios Kotsopoulos\textsuperscript{2},
Francesca Dominici\textsuperscript{3}}
\IEEEauthorblockA{\textsuperscript{1}\textit{TWT GmbH Science \& Innovation}, Stuttgart, Germany}
\IEEEauthorblockA{\textsuperscript{2}\textit{National Technical University of Athens}, Athens, Greece}
\IEEEauthorblockA{\textsuperscript{3}\textit{Harvard University}, USA}
\IEEEauthorblockA{\textsuperscript{4}\textit{Massachusetts Institute of Technology}, Cambridge, MA, USA}
\IEEEauthorblockA{\textsuperscript{5}\textit{NIKI Digital Engineering}}
\IEEEauthorblockA{\textsuperscript{6}\textit{Technical University of Munich}, Munich, Germany}
\IEEEauthorblockA{\textsuperscript{7}\textit{Siemens Research}, Germany}
\IEEEauthorblockA{tina.vartziotis@twt-gmbh.de, ketevan.peranidze@twt-gmbh.de, gdasoulas@hsph.harvard.edu}
}

\maketitle

\begin{abstract}
As large language models (LLMs) evolve from conversational assistants into
autonomous agents embedded in software development workflows, understanding
their environmental footprint requires moving beyond conventional token-based
accounting. Despite their growing adoption, we still lack empirical evidence
of how these systems consume computational resources in real-world workflows.
Existing energy estimators largely focus on isolated request--response
interactions, overlooking the repeated processing of accumulated context,
cache reuse, and computational work delegated to subagents. We address this
gap by analyzing nearly one billion tokens collected across 42 sessions and
752 task episodes over ten weeks of real-world AI-assisted software development.
Our analysis reveals that \num{94.8}\% of all API-reported tokens are
cached-context reads, while uncached input accounts for only \num{0.017}\%.
For every generated token, approximately 260 cached tokens are read.
Context grows 5.3-fold over a session compared with 1.4-fold within
individual tasks, revealing how computational demands accumulate over
long-running workflows. To quantify the energy implications, we extend an
existing analytical inference-energy model to account for cached-context
processing, yielding a cache-aware hybrid estimator. It estimates
approximately 250~kWh for the workload, compared with approximately 10~kWh
from the output-based estimator EcoLogits, a 25-fold difference largely
driven by cached-context reads. Neither estimator has been validated on long
cached contexts, but no modeling assumption we varied closes the gap.
Starting each task from a fresh context would lower our estimate by about
60\%. We also find that
parent-session logs report only a subagent's final internal call, undercounting
delegated tokens by a factor of ten; for this workload the correction changes
the energy estimate by about 5\%, but token-volume studies are fully exposed
to it. Our findings expose a substantial gap between how current energy
estimators and agentic workloads treat context, highlighting
the need for workflow-level energy accounting that captures context
accumulation, cache reuse, and delegated execution.
\end{abstract}


\input{sections/introduction}
\input{sections/corpus}
\section{Results}\label{sec:resultsall}

\input{sections/workload}
\input{sections/energy}
\input{sections/subagents}
\input{sections/discussion}
\input{sections/conclusion}

\section*{Acknowledgment}
Claude Code, the assistant studied in this paper, was used to write the
analysis and experiment code and to assist in editing the text. All results
were checked against the underlying data by the authors.

\bibliographystyle{IEEEtran}
\bibliography{paper_resources/references}

\clearpage
\input{sections/appendix}

\end{document}

%% file: sections/introduction.tex
\section{Introduction}

Agentic AI assistants are increasingly transforming software engineering,
enabling autonomous repository exploration, code modification, testing, and
complex development workflows~\cite{yang2024sweagent,sarikayak2026swepro}.
Unlike conventional question--answer interactions, these systems operate
through sequences of model invocations, tool executions, and delegated
subtasks. As their adoption grows, understanding the computational resources
required to sustain these workflows becomes increasingly important for
assessing the environmental footprint of AI-assisted software
engineering~\cite{verdecchia2021green}.

Measuring the resource demands of agentic AI presents challenges
that extend beyond conventional LLM inference. Each model invocation may
reuse accumulated conversation history, retrieved code, and tool outputs,
causing the processed context to grow throughout a development session.
Although prompt caching avoids redundant prefill
computation~\cite{gim2024promptcache,anthropic2026caching}, cached context
can still contribute to attention computation and memory-access costs during
inference~\cite{kwon2023pagedattention}. Moreover, agentic workflows may
delegate tasks to subagents whose internal model invocations are not fully
visible in parent-level execution records. These characteristics complicate
the relationship between reported token counts, actual inference operations,
and estimated energy consumption.

Existing research has established important foundations for assessing LLM
inference energy. Measurement-based studies examine the influence of model
architecture, hardware, and inference configurations on energy
consumption~\cite{samsi2023words,luccioni2024power,fernandez2025energy},
while analytical approaches estimate resource demands from token counts and
model-level computational characteristics~\cite{ecologits2025,vartziotis2026tokens}.
However, these approaches do not necessarily capture the characteristics
of sustained agentic execution. In particular, estimators that do not
explicitly distinguish cached from freshly processed context can overlook
important inference operations or assign inappropriate computational costs
to reused tokens. Similarly, analyses relying on parent-level usage records
may miss computation performed within delegated execution.

Despite the growing deployment of coding agents, empirical evidence of
their resource consumption across real-world development workflows remains
limited. The distribution of cached reads, cache writes, uncached input,
and generated output, as well as the evolution of these components across
sessions and individual tasks is not yet sufficiently characterized in
energy studies. Consequently, it remains unclear how these workload
characteristics affect energy estimates and to what extent incomplete
execution traces distort resource accounting.

To address these gaps, we study ten weeks of real-world Claude Code use in an
industrial research environment: 42 sessions and 752 task episodes across six
workstreams, totaling \num{0.99} billion cumulative API-reported tokens. Our
contributions are threefold. (1)~\emph{An empirical characterization of
agentic workloads:} cached-context reads make up \SI{94.8}{\percent} of
tokens, and context grows 5.3-fold over a session but only 1.4-fold within a
task. (2)~\emph{Cache-aware energy estimation:} a hybrid estimator that
separates cached from freshly processed tokens assigns the workload about 25
times the energy EcoLogits does, a gap that persists under every modeling
assumption we varied; we also show that on this trace starting each task from
a fresh context would lower the model-based estimate by about
\SI{60}{\percent}. (3)~\emph{Reconstruction of delegated workloads:}
parent-level records undercount subagent tokens tenfold, because only the
final internal call is reported.

%% file: sections/corpus.tex
\section{Methodology}\label{sec:methodology}

\begin{figure*}[t]
\centering
\footnotesize
\begin{tikzpicture}[
    box/.style    = {draw, rounded corners=1.5pt, align=left,
                     inner sep=4pt, text width=#1},
    shaded/.style = {box=#1, fill=black!8},
    arr/.style    = {-{Stealth[length=1.6mm]}, semithick},
  ]

\node[box=26mm, anchor=west] (src) at (0,0) {%
  \textbf{Session transcripts}\\[2pt]
  45 JSONL transcripts\\42 sessions\\
  6 workstreams\\9\,Jun–18\,Aug\,2026};

\node[box=34mm, anchor=west] (dedup) at ($(src.east)+(4mm,0)$) {%
  \textbf{Deduplication}\\[2pt]
  1.\ resumed-session copies\\
  \quad removed (\num{752} copies\\
  \quad of \num{607} responses)\\
  2.\ \num{7706} records merged\\
  \quad into \num{3738} responses\\
  3.\ \num{15} synthetic removed};

\node[box=24mm, anchor=west] (calls) at ($(dedup.east)+(4mm,0)$) {%
  \textbf{\num{3723} API calls}\\[2pt]
  Four token classes:\\[1pt]
  $T_{\mathrm{in}},\;T_{\mathrm{write}},$\\
  $T_{\mathrm{read}},\;T_{\mathrm{out}}$};

\node[box=42mm, anchor=west] (work) at ($(calls.east)+(8mm,14mm)$) {%
  \textbf{Workload characterization}\\[2pt]
  752 episodes: declared task type,\\
  tool-use label, context by\\
  progress decile};

\node[shaded=42mm, anchor=west] (energy) at ($(calls.east)+(8mm,-18mm)$) {%
  \textbf{Energy, per request}\\[2pt]
  EcoLogits (reference): $T_{\mathrm{out}}$ only\\[1pt]
  Hybrid, three cache scenarios:\\
  \quad omitted: $T_{\mathrm{read}}$ dropped\\
  \quad cache-aware: $T_{\mathrm{read}}$ attended only\\
  \quad cache-naive: $T_{\mathrm{read}}$ as fresh\\[1pt]
  Both: + server, PUE, embodied, grid};

\node[shaded=19mm, anchor=west] (checks) at ($(energy.east)+(5mm,0)$) {%
  \textbf{Checks}\\[2pt]
  one-factor\\sensitivity\\[1pt]
  full-power\\energy per bit\\[1pt]
  decoding-time\\experiment};

\node[box=62mm, anchor=north west] (deleg) at ($(src.south west)+(0,-5mm)$) {%
  \textbf{123 delegations:} 106 foreground,
  whose result reports only the final internal
  call $\to$\;\num{1334} internal calls from
  private transcripts; 17 background, reporting
  none $\to$ 8 recovered (\num{104} calls)};

\draw[arr] (src.east)   -- (dedup.west);
\draw[arr] (dedup.east) -- (calls.west);

\coordinate (fk) at ($(calls.east)+(4mm,0)$);
\draw[semithick] (calls.east) -- (fk);
\draw[arr] (fk) |- (work.west);
\draw[arr] (fk) |- ($(energy.west)+(0,8mm)$);

\draw[arr] (deleg.east) -- (energy.west |- deleg.east);

\draw[arr] (energy.east) -- (checks.west);

\node[font=\scriptsize\itshape, text=black!60, text width=155mm,
      anchor=north west, inner sep=0pt]
  at ($(deleg.west |- energy.south)+(0,-2mm)$) {%
  Boundaries: main-thread trace (\num{3723} calls)
  $\cdot$ accounted workload (+106 delegations as reported $=$ \num{3829})
  $\cdot$ reconstructed workload (106 replaced by \num{1334} internal calls,
  +\num{104} background $=$ \num{5161}).};

\end{tikzpicture}
\caption{Study pipeline. Usage is recorded in the transcripts; API calls and
delegated runs are reconstructed from them; energy is modeled per request by
two estimators, the hybrid one in three cache scenarios. Tokens are reported
for the main-thread trace, energy for the accounted and reconstructed
workloads. Shaded: model-based.}
\label{fig:method}
\end{figure*}
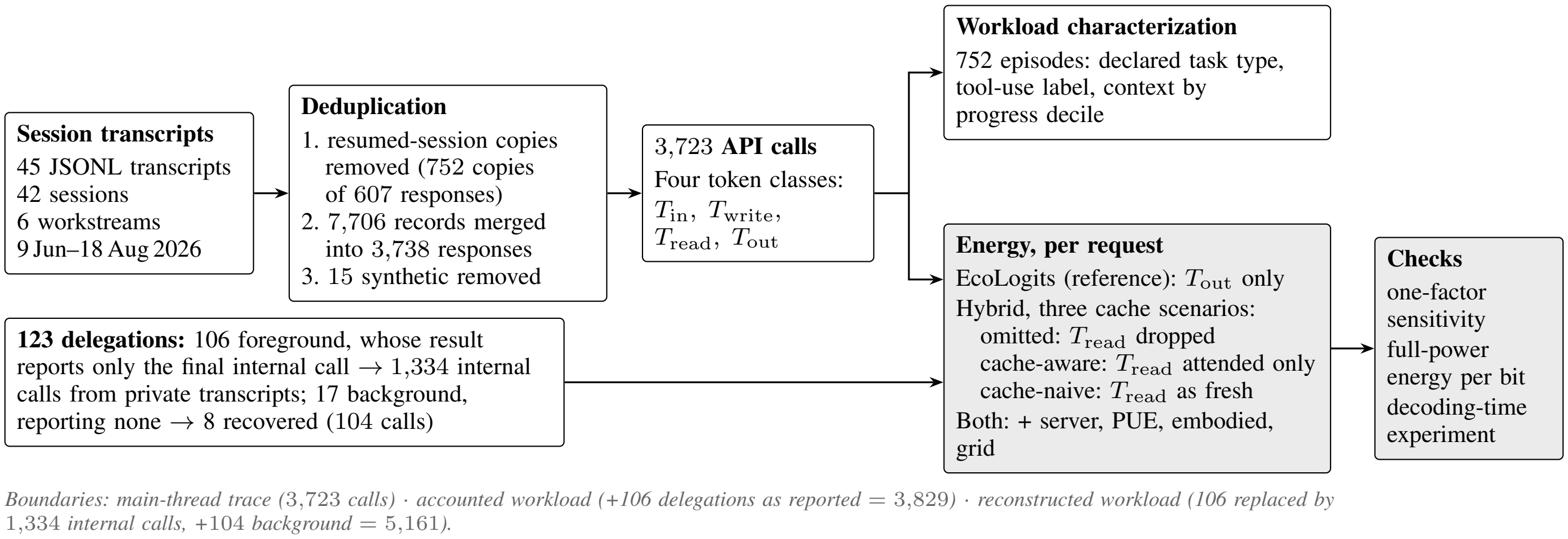

\subsection{Study Design and Data}\label{sec:study-design}\label{sec:corpus}

We conducted a retrospective observational study of API-reported token usage
in routine agentic software engineering (Fig.~\ref{fig:method}). The trace
covers the use of Claude Code~\cite{anthropic2026claudecode} by a single
engineer between 9 June and 18 August 2026 across six workstreams: three software products, OPTIMALIS~\cite{twt2025optimalis}
(workstream \wsPA), Artemis and FORMALIS~\cite{twt2025formalis} (\wsPC); two
research projects, GreenCode~\cite{itea2024greencode,greencode2026web}
(\wsRA) and \wsRB; and one customer project that is not named for reasons of
confidentiality.{} It contains seven model versions from three
tiers (Haiku~4.5; Sonnet~4.6 and~5; Opus~4.6--4.8 and~5), selected as in
ordinary use. Token counts, timestamps, model identifiers and tool invocations
are recorded; energy and emissions are modeled from them, without measurements
on the vendor's hardware. Token totals are cumulative over requests: a cached
prefix is counted again by every request that reads it.

A \emph{session} is one Claude Code conversation, stored as a transcript
file; a resumed session continues an earlier one in a new file and is merged
with it. The \emph{main thread} is the conversation between engineer and
assistant, which can \emph{delegate} a subtask to a \emph{subagent}, a separate
assistant instance with its own context and transcript. A \emph{foreground}
delegation blocks the main thread until it returns; a \emph{background}
delegation runs asynchronously.

Table~\ref{tab:corpus} summarizes the trace by workstream. It comprises the
engineer's project directories of more than about one million tokens,
selected retrospectively by inspection (Appendix~\ref{app:data}). The derived tables and the analysis code are
available from the authors upon request; the
transcripts are withheld because they contain proprietary code.

\begin{table}[t]
\centering
\caption{The ten-week trace by workstream. Transcr.: transcript files, three
of which resume earlier sessions (42 sessions after merging). A turn is one
API response, not one transcript record (Section~\ref{sec:processing}).}
\label{tab:corpus}
\footnotesize
\begin{tabular}{lrrrr}
\toprule
Workstream & Transcr. & Turns & Ep. & Tokens \\
\midrule
\wsPA{}         &  9 & \num{1117} & \num{188} & \num{474726571} \\
\wsRA{}        & 12 &  \num{916} & \num{209} & \num{332447525} \\
\wsRB{}        &  9 &  \num{646} & \num{139} & \num{85182813} \\
\wsCu{}          &  6 &  \num{596} & \num{106} & \num{56681332} \\
\wsPB{}         &  7 &  \num{405} &  \num{89} & \num{40935105} \\
\wsPC{}         &  2 &   \num{58} &  \num{21} & \num{2575698} \\
\midrule
Total             & 45 & \num{3738} & \num{752} & \num{992549044} \\
\bottomrule
\end{tabular}
\end{table}

\subsection{API Calls, Episodes and Delegations}\label{sec:processing}

The transcripts write one record per content block of a response, each with a
copy of the response's usage, and a resumed session begins with a copy of the
session it continues. We remove the copies and count each response, hereafter
a \emph{turn}, once: \num{3738} turns, of which \num{3723} are API calls; the
others are synthetic records written by the client (Appendix~\ref{app:collapse}).

For each call $i$ the API reports four token counts: uncached input
($T_{i,\mathrm{in}}$), cache writes ($T_{i,\mathrm{write}}$), cache reads
($T_{i,\mathrm{read}}$) and output ($T_{i,\mathrm{out}}$). Cache writes are new
input that is processed and then stored, so the model processes
$T_{\mathrm{fresh}}$ tokens anew, reuses $T_{\mathrm{cached}}$ tokens from the
cache, and attends over a context of $T_{\mathrm{ctx}}$ tokens (index $i$
omitted):
\begin{gather*}
T_{\mathrm{fresh}} = T_{\mathrm{in}} + T_{\mathrm{write}}, \qquad
T_{\mathrm{cached}} = T_{\mathrm{read}}, \\
T_{\mathrm{ctx}} = T_{\mathrm{fresh}} + T_{\mathrm{cached}}.
\end{gather*}

An \emph{episode} is one task request, a human prompt together with any
further prompts sent before the assistant responds, and all assistant calls
until the next request. The 994 prompts form 752 episodes, 746 of them with
at least one call (\emph{working episodes}). Each episode carries a
\emph{declared} task type from keyword rules and a \emph{behavioral} label,
its most frequent tool category (Appendix~\ref{app:data}).

\subsubsection*{Delegations}\label{sec:delegations}\label{sec:evaluation}
The assistant delegated 123 subtasks (106 foreground, 17 background) to
subagents, whose own calls are
recorded in separate transcripts. Token results describe the
\emph{main-thread trace} (\num{3723} calls). Energy results use the
\emph{accounted workload} (\num{3829} requests), which adds each delegation
with the usage reported to the main thread, or the \emph{reconstructed
workload} (\num{5161} requests), which replaces those reports with the
subagents' own calls (Section~\ref{sec:subagents}).

\subsection{Energy and Carbon Estimation}\label{sec:energy-framework}

We compare two estimators (Table~\ref{tab:estimators}). The reference is
EcoLogits v0.11.1~\cite{ecologits2025}. The \emph{hybrid} estimator combines
the analytical inference-energy model of Vartziotis
\emph{et al.}~\cite{vartziotis2026tokens} with the infrastructure methodology
of EcoLogits and extends it to prompt caching. To isolate the effect of
representing cache reuse, the hybrid estimator is evaluated in three cache
scenarios that differ only in how cache reads are charged; hybrid results
without a named scenario are cache-aware. Of the equations
below, Eqs.~\eqref{eq:eco}, \eqref{eq:facility} and~\eqref{eq:carbon} are
taken from EcoLogits; Eqs.~\eqref{eq:hybrid} and~\eqref{eq:attn} are terms of
the analytical model, into which our extension enters different token counts
and key--value width; Eq.~\eqref{eq:bounds} is derived here.

\begin{table}[t]
\centering
\caption{How each estimator charges the token classes of a request, and the
facility energy it assigns to a typical request: the main-thread call closest
to the medians of context, fresh input and output (804 fresh, \num{153355}
cached and 389 output tokens on Sonnet~4.6; central values). P: dense prefill and KV write; D: decode (hybrid: including KV
write); A: read by attention; ---: not charged. All variants add server energy, PUE and
embodied impact (Eqs.~\eqref{eq:facility}--\eqref{eq:carbon}).}
\label{tab:estimators}
\footnotesize
\setlength{\tabcolsep}{4pt}
\begin{tabular}{@{}lcccr@{}}
\toprule
Estimator & $T_{\mathrm{fresh}}$ & $T_{\mathrm{cached}}$ & $T_{\mathrm{out}}$
  & Typical (\si{\watt\hour}) \\
\midrule
EcoLogits (reference)  & ---  & ---  & D    & 0.71 \\
Hybrid, cache omitted  & P, A & ---  & D, A & 0.66 \\
Hybrid, cache-aware    & P, A & A    & D, A & 11.5 \\
Hybrid, cache-naive    & P, A & P, A & D, A & 30.1 \\
\bottomrule
\end{tabular}
\end{table}

Fresh tokens pass through the full forward computation (prefill). Cached
tokens skip it, because their key--value entries are stored, but every
generated token still reads those entries in attention, at a cost
proportional to their number, much as rereading one's notes is cheaper than a
first reading but not free. EcoLogits ignores input tokens altogether: its GPU
energy depends on output tokens only, and its server and embodied shares on
request latency, which we take from the transcript timestamps. The
cache-omitted scenario charges fresh input and output but treats cached tokens
as free; the cache-aware scenario charges cached tokens for attention reads
only; the cache-naive scenario charges them as fresh input. For the typical
request of Table~\ref{tab:estimators}, re-reading is therefore neither free
nor as costly as a first reading: its cached tokens cost \SI{3.7}{\watt\hour}
of attention reads, whereas processing them as fresh input would add
\SI{9.3}{\watt\hour} of prefill (GPU energy, low bound).

In the EcoLogits methodology~\cite{ecologits2025}, the GPU energy of request
$i$ is
\begin{equation}
E_{i,\mathrm{GPU}}^{\mathrm{Eco}} = T_{i,\mathrm{out}} \cdot f_E(N_i, B),
\label{eq:eco}
\end{equation}
where $f_E$ is its fitted energy per output token for active model size $N_i$
and batch size $B$.

\subsubsection*{Cache-aware GPU estimator}\label{sec:cache-model}\label{sec:hybrid}
The analytical model represents prefill and decode computation, parameter
accesses, key--value (KV) writes and attention reads
(Appendix~\ref{app:original}). Its energy coefficients come from
microbenchmarks on the Hopper GPU of an NVIDIA
GH200~\cite{antepara2025benchmark}, and its memory factors were fitted to dense
open-weight models of 8B to 72B parameters~\cite{caravaca2025prompts}, which it
reproduces within \SIrange{5}{27}{\percent} for uncached requests of 500 input
and 500 output tokens. The validation covers uncached requests only; the
contexts of this trace are two orders of magnitude longer, so the
attention-read term that dominates our estimates is evaluated outside the
validated range.

It writes the GPU energy of a request as the sum of five
terms~\cite{vartziotis2026tokens},
\begin{equation}\label{eq:hybrid}
E^{\mathrm{hyb}}_{i,\mathrm{GPU}}=
E_{i,\mathrm{prefill}}+E_{i,\mathrm{decode}}+
E_{i,\mathrm{attn}}+E_{i,\mathrm{KV}}+E_{i,\mathrm{param}}.
\end{equation}
The original model treats all input as an uncached prompt. Our extension
changes which token counts enter the terms: fresh tokens require forward
computation, so dense prefill is evaluated over $T_{i,\mathrm{fresh}}$;
cached tokens bypass it but remain available to attention, so the attention
terms are evaluated over the full context $T_{i,\mathrm{ctx}}$
(Table~\ref{tab:app-extension}, Appendix~\ref{app:extension}, compares every
term). Decoding
$T_{i,\mathrm{out}}$ tokens reads
$A_i=T_{i,\mathrm{out}}T_{i,\mathrm{ctx}}+T_{i,\mathrm{out}}(T_{i,\mathrm{out}}-1)/2$
context positions, and the memory energy of these attention reads is
\begin{equation}\label{eq:attn}
E_{i,\mathrm{attn}}=e_{\mathrm{mem}}\,\eta(N_i)\,
 s_{\mathrm{attn}}(N_i)\,b_{\mathrm{KV},i}\,A_i,
\end{equation}
with memory energy per bit $e_{\mathrm{mem}}$, the memory-inefficiency and
attention-read factors $\eta$ and $s_{\mathrm{attn}}$ calibrated in the
original model, and KV-entry size $b_{\mathrm{KV},i}=2n_{\mathrm{layers},i}d_{kv}b$
bits for $b$-bit values. Eq.~\eqref{eq:attn} is the attention-read term
of~\cite{vartziotis2026tokens}, which we evaluate over $T_{i,\mathrm{ctx}}$
instead of the uncached prompt. The entry size uses the effective key--value
dimension $d_{kv}$ of grouped-query attention, to which the original text
refers, where its equations use the hidden dimension $d_{\mathrm{model}}$
(Appendix~\ref{app:extension}). Because
Eq.~\eqref{eq:attn} depends on the \emph{product} of context and output, the
estimator is applied per request and then summed; no coefficient is refitted
to the trace. Writing $E_{\mathrm{GPU}}^{\mathrm{hyb}}(T_{\mathrm{fresh}},
T_{\mathrm{cached}}, T_{\mathrm{out}})$ for the estimate as a function of its
token inputs, we show that, since every term is non-decreasing in its token
counts, the three cache scenarios are ordered:
\begin{multline}
E_{\mathrm{GPU}}^{\mathrm{hyb}}(T_{i,\mathrm{fresh}}, 0, T_{i,\mathrm{out}})
\le E_{\mathrm{GPU}}^{\mathrm{hyb}}(T_{i,\mathrm{fresh}}, T_{i,\mathrm{cached}}, T_{i,\mathrm{out}}) \\
\le E_{\mathrm{GPU}}^{\mathrm{hyb}}(T_{i,\mathrm{ctx}}, 0, T_{i,\mathrm{out}}),
\label{eq:bounds}
\end{multline}
cache omitted, cache-aware and cache-naive (proof in
Appendix~\ref{app:bounds}). The ordering follows from the structure of the
model and holds for any values of its coefficients.

\subsubsection*{Facility energy and emissions}\label{sec:facility-carbon}\label{ssec:carbon}
Both remaining steps follow the EcoLogits methodology~\cite{ecologits2025},
with the PUE and batch size of Section~\ref{sec:assumptions}
(Appendix~\ref{app:layer2}). Non-GPU server energy is added and power usage
effectiveness (PUE) applied,
\begin{equation}\label{eq:facility}
E^{\mathrm{hyb}}_{i,\mathrm{facility}}=
\mathrm{PUE}\bigl(E^{\mathrm{hyb}}_{i,\mathrm{GPU}}+E_{i,\mathrm{server}}\bigr),
\end{equation}
where the bracketed sum is \emph{IT energy}, and emissions are
\begin{equation}\label{eq:carbon}
\mathrm{GWP}_i=I_{\mathrm{grid}}E^{\mathrm{hyb}}_{i,\mathrm{facility}}+
\mathrm{GWP}_{i,\mathrm{emb}},
\end{equation}
with allocated embodied impact $\mathrm{GWP}_{i,\mathrm{emb}}$. The US-average
$I_{\mathrm{grid}} = \SI{0.3844}{\kilo\gram\,CO_2eq\per\kilo\watt\hour}$ used
by EcoLogits~\cite{owid2026carbon} is the baseline; Section~\ref{ssec:energy-total}
evaluates other grids. The boundary excludes client-side tools, network
transfer, training, retention of cached entries between requests, and idle
serving capacity.

\subsection{Assumptions and Analyses}\label{sec:assumptions}\label{ssec:analyses}

As the models are proprietary, their architecture is derived from the lower
and upper active-parameter estimates of the EcoLogits registry
(Appendix~\ref{app:estimator}). Every request is evaluated at both bounds, and
the central estimate is their geometric mean, a scenario range rather than a
confidence interval. Headline totals are central values; comparisons between
workloads, the counterfactuals and the sensitivity analyses use the lower
bound. The hybrid model uses $B = 16$ and $\mathrm{PUE} = 1.2$, EcoLogits its
own defaults.

Context accumulation is measured over deciles of turn position in sessions
with at least 20 turns and episodes with at least 10. Two equal-trace
counterfactuals re-estimate the recorded requests with everything else held
fixed: the \emph{model} counterfactual runs all of them on each of five models
(Haiku~4.5, Sonnet~4.6, Opus~4.8, GPT-5.4, Gemini~2.5~Pro), and the
\emph{fresh-context} counterfactual starts every episode in a new session
(Appendix~\ref{app:fresh}). Sensitivity is assessed by varying each assumption
one at a time (Table~\ref{tab:factors}), and a decoding-time experiment on
three open-weight models tests the linear attention term of
Eq.~\eqref{eq:attn} (Appendix~\ref{app:sweep}).

%% file: sections/workload.tex
\subsection{Characterization of the Agentic Workload}\label{sec:results}

\subsubsection{Token Composition}

\begin{table}[t]
\centering
\caption{Token composition of the main-thread trace.}
\label{tab:composition}
\footnotesize
\begin{tabular}{lrr}
\toprule
Class & Tokens & Share \\
\midrule
Cache read   & \num{940461095}  & \SI{94.752}{\percent} \\
Cache write  & \num{48302053}   & \SI{4.866}{\percent} \\
Output       & \num{3618355}    & \SI{0.365}{\percent} \\
Uncached input & \num{167541}     & \SI{0.017}{\percent} \\
\midrule
Total        & \num{992549044}  & \SI{100}{\percent} \\
\bottomrule
\end{tabular}
\end{table}

Cache reads account for \SI{94.75}{\percent} of all tokens and output for
\SI{0.36}{\percent} (Table~\ref{tab:composition}), i.e., \num{260} cached tokens
are read per token generated. Uncached input $T_{\mathrm{in}}$, the only input
that the API's input-token field reports, accounts for \SI{0.017}{\percent}.
An estimator based on $T_{\mathrm{in}}$ and $T_{\mathrm{out}}$ therefore sees
\SI{0.38}{\percent} of the recorded tokens, whereas treating all tokens as
fresh input would charge full prefill to \num{0.94} billion tokens served from
the cache. The composition is similar across workstreams, from
\SI{86.6}{\percent} cache reads in \wsPC{} to \SI{95.7}{\percent} in
\wsPA{} (Fig.~\ref{fig:composition}, Appendix~\ref{app:workload}).
These are the token classes reported and billed by the API, not observations
of how the serving hardware processed them; any energy estimate for this
workload has to state how it treats cached tokens.

\subsubsection{Context Accumulation}

\begin{figure}[t]
\centering
\includegraphics[width=\columnwidth]{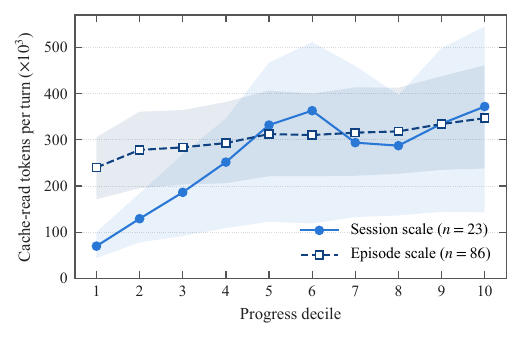}
\caption{Mean cache-read tokens per turn by decile of progress, at session scale
(sessions of at least \num{20} turns) and episode scale (episodes of at least
\num{10} turns). Means are pooled over all turns in a decile, so longer
sessions contribute more turns. Shaded bands are \SI{95}{\percent} confidence
intervals from a cluster bootstrap over sessions and episodes, respectively
(\num{2000} resamples).}
\label{fig:context}
\end{figure}

Over a session (\num{23} sessions with at least \num{20} turns, a resumed
session counted with the one it continues), mean cache-read tokens per turn
grow from \num{70}k in the first decile to \num{372}k in the last, a factor of
\num{5.3} (Fig.~\ref{fig:context}); within an episode (\num{86} episodes with
at least \num{10} turns), they grow only by \num{1.4}. Weighting sessions
equally gives \num{3.8} and \num{1.3}, and context grows in all \num{23}
sessions. Output per turn stays roughly constant over a session (factor
\num{0.9}) and falls within an episode (\num{0.3}), so later turns read more
and generate less. Episodes thus inherit most of their context from earlier
work in the session, and resource accounting has to follow context across
tasks rather than treat each task as independent. The analysis does not
separate session length from task complexity or context compaction (the
automatic summarization of a context that approaches its limit).

\subsubsection{Token Cost by Task Type}

\begin{figure*}[t]
\centering
\includegraphics[width=\textwidth]{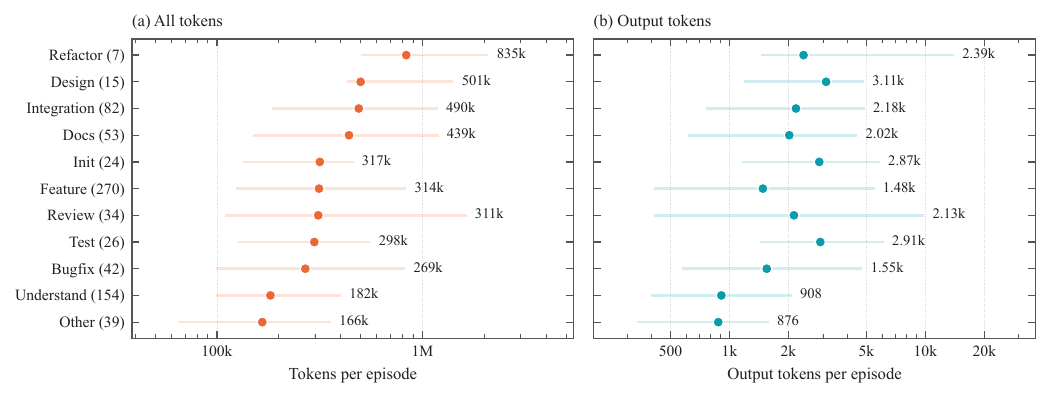}
\caption{Median total tokens (a) and median output tokens (b) per episode by
declared task type, working episodes only, on logarithmic axes; faint bars span the interquartile range. The number of
episodes is given in parentheses.}
\label{fig:tasks}
\end{figure*}

Median episode cost ranges from \num{166}k tokens for Other to \num{835}k for
Refactor (Fig.~\ref{fig:tasks}; Table~\ref{tab:tasks},
Appendix~\ref{app:workload}). Output ranks differently: Design, Test and Init
generate the most (medians of \num{2.9}k to \num{3.1}k tokens), whereas
Refactor ranks fourth and Integration, third in total, fifth. The differences
mostly reflect episode length and session position: in a descriptive
regression of log episode tokens, turn count, session position and primary
model explain \SI{88.9}{\percent} of the variance, the task label alone
\SI{6.3}{\percent}, and both together \SI{89.3}{\percent}
(Appendix~\ref{app:workload}). As the labels come from unvalidated keyword
rules and Refactor (\num{7} episodes) and Design (\num{15}) are small samples,
the medians describe this workload rather than task types in general. Tool use
does not explain cost either: \SI{46}{\percent} of working episodes, and
\SI{73}{\percent} of Understand episodes, invoke no tool, yet Understand
episodes consume a median of \num{182}k tokens because the context is re-read
regardless.

%% file: sections/energy.tex
\subsection{Energy Implications of Cached Context}
\label{sec:energy}

\subsubsection{Total Energy and Carbon}
\label{ssec:energy-total}

For the \num{3829} requests of the accounted workload, the hybrid estimate
exceeds the EcoLogits estimate by a factor of about \num{25} in energy and
\num{24} in carbon emissions (Table~\ref{tab:energy-headline}). The ratio
compares two models, EcoLogits~\cite{ecologits2025} and the hybrid extension
of~\cite{vartziotis2026tokens}. Neither has been validated against metered
energy for requests with long cached contexts such as those in this trace, so
the ratio quantifies the disagreement between the two models. They differ in
their treatment of context, their GPU energy model and, to a minor extent,
batch size and PUE, which EcoLogits fixes internally
(Appendix~\ref{app:estimator}). The disagreement is structural: EcoLogits has
no input that responds to context, whereas cache reads make up
\SI{95}{\percent} of the tokens here. Its magnitude depends on the modeling
assumptions; its sign does not, as no assumption we varied closes the gap
(Section~\ref{ssec:energy-sensitivity}). The scenarios of Eq.~\eqref{eq:bounds} isolate the cached context: omitting it
yields \SI{36.8}{\kilo\watt\hour}, so accounting for it contributes a factor of
\num{6.7} to the ratio and the remaining differences about \num{3.7}. The
cache-naive variant is \num{1.8} times the hybrid estimate at the central
values (\num{2.2} at the lower bound, \num{1.5} at the upper), so prompt caching
reduces the estimated energy by about \SI{46}{\percent}. Unless stated
otherwise, the comparisons below and in Section~\ref{sec:subagents} use the
lower bound (\SI{105.0}{\kilo\watt\hour}), at which the hybrid/EcoLogits
ratio is \num{14.5} rather than \num{25}.

\paragraph*{Scale}
In familiar terms (Table~\ref{tab:scale}), the estimator decides where
agentic use stands: under EcoLogits a median request costs as much as three
to four chatbot prompts, under the hybrid estimate as much as 60 to 77.

\paragraph*{Carbon and the grid}
Emissions apply one grid factor to all requests, the 2025 lifecycle carbon
intensity of US electricity (\SI{384}{\gram} CO$_2$eq/\si{\kilo\watt\hour})
that EcoLogits takes from Our World in Data~\cite{owid2026carbon}. It is a location-based annual national average;
it reflects neither the serving region, which is not
recorded~\cite{anthropic2026residency}, nor renewable procurement such as
power purchase agreements, which a market-based account would
include~\cite{ghgp2015scope2}. Because usage emissions scale with the grid
factor, the central hybrid estimate ranges from \SI{9.7}{\kilo\gram} on the
Swedish grid (\SI{35}{\gram}/\si{\kilo\watt\hour}) and \SI{71}{\kilo\gram} at
the European average (\SI{283}{\gram}) to \SI{147}{\kilo\gram} on the
coal-dominated Polish grid (\SI{589}{\gram}); EcoLogits gives \num{0.5},
\num{3.0} and \SI{6.0}{\kilo\gram}. Average factors attribute to the workload the
emissions of the existing mix; for an incremental workload, marginal factors,
which differ substantially by region and hour~\cite{silerevans2012mef}, would
be the better basis for decisions but require the serving location and time.

\begin{table}[t]
\centering
\caption{Modeled totals for the accounted workload of \num{3829} API calls
(\num{3723} main-thread, \num{106} delegated), US grid. Central value: geometric
mean of the two active-parameter bounds, given in brackets as low--high. The hybrid rows are the three scenarios
of Eq.~\eqref{eq:bounds}: cache omitted, cache-aware, and cache-naive
(the full context as fresh input).}
\label{tab:energy-headline}
\footnotesize
\begin{tabular}{lrr}
\toprule
 & Energy (\si{\kilo\watt\hour}) & Carbon (\si{\kilo\gram} CO$_2$eq) \\
\midrule
EcoLogits & 9.89 & 3.97 \\
          & [7.24--13.5] & [2.94--5.35] \\
Cache omitted & 36.8 & 15.0 \\
          & [18.5--73.4] & [7.82--29.0] \\
Cache-aware & 248 & 96.2 \\
          & [105--585] & [41.1--225] \\
Cache-naive & 456 & 176 \\
          & [232--895] & [90.0--345] \\
\midrule
Ratio cache-aware/EcoLogits & 25.1$\times$ & 24.2$\times$ \\
\midrule
Per call, EcoLogits & \SI{2.58}{\watt\hour} & \SI{1.04}{\gram} \\
Per call, cache-aware & \SI{64.7}{\watt\hour} & \SI{25.1}{\gram} \\
\bottomrule
\end{tabular}
\end{table}

\begin{table}[t]
\centering
\caption{The workload in familiar terms, central values (hybrid at the low
bound: \num{2.4} times smaller). Engineer-year: 220 working days. Household:
\SI{863}{\kilo\watt\hour} a month, US 2024~\cite{eia2025sales}. Car:
\SI{400}{\gram} tailpipe CO$_2$ per mile~\cite{epa2026vehicle}. Chatbot
prompt: median \SIrange{0.24}{0.31}{\watt\hour}~\cite{elsworth2025google,oviedo2026joule}.
Median request: median per-call energy (\num{0.91} and \SI{18.5}{\watt\hour}),
not the typical request of Table~\ref{tab:estimators}.}
\label{tab:scale}
\footnotesize
\setlength{\tabcolsep}{4pt}
\begin{tabular}{@{}lrr@{}}
\toprule
 & EcoLogits & Hybrid \\
\midrule
\multicolumn{3}{@{}l}{\emph{Per engineer}} \\
Electricity per working day          & \SI{0.19}{\kilo\watt\hour} & \SI{4.9}{\kilo\watt\hour} \\
Electricity per engineer-year        & \SI{43}{\kilo\watt\hour}   & \SI{1070}{\kilo\watt\hour} \\
\quad share of a US household's year & \SI{0.4}{\percent}         & \SI{10}{\percent} \\
Ten weeks of emissions, in car travel & \SI{16}{\kilo\meter}      & \SI{390}{\kilo\meter} \\
\midrule
\multicolumn{3}{@{}l}{\emph{Scaled, per year}} \\
Team of ten engineers                & \SI{0.43}{\mega\watt\hour} & \SI{11}{\mega\watt\hour} \\
Per million engineers                & \SI{0.04}{\tera\watt\hour} & \SI{1.1}{\tera\watt\hour} \\
\midrule
\multicolumn{3}{@{}l}{\emph{Compared with}} \\
Median request / chatbot prompt      & 3--4$\times$               & 60--77$\times$ \\
\bottomrule
\end{tabular}
\end{table}

\subsubsection{Decomposition by Physical Term}

\begin{figure}[t]
\centering
\includegraphics[width=\columnwidth]{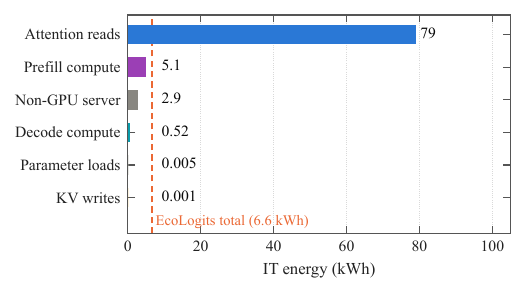}
\caption{Hybrid IT energy by physical term, low active-parameter bound. The
dashed line marks the total EcoLogits estimate for the same trace on the same
basis (IT energy, low bound, \SI{6.6}{\kilo\watt\hour}).}
\label{fig:e2}
\end{figure}

Within the hybrid model, the difference arises from a single term
(Fig.~\ref{fig:e2}; Table~\ref{tab:energy-terms}, Appendix~\ref{app:energy}):
attention reads over the full context during decoding make up about
\SI{90}{\percent} of IT energy, \num{12} times the complete EcoLogits estimate
on the same basis (\SI{6.6}{\kilo\watt\hour}). This is the term that an
output-only estimator lacks, and its size depends on the assumed key--value
representation (Section~\ref{ssec:energy-sensitivity}). Per-call hybrid
energy therefore follows the context (Spearman $\rho = 0.59$ with
$T_{\mathrm{ctx}}$, $0.34$ with uncached input) and, at fixed output, spreads
by up to a factor of \num{14} where EcoLogits spreads by \num{1.4}
(Appendix~\ref{ssec:energy-vs-output}).

\subsection{Model Choice and Modeling Assumptions}
\label{sec:sensitivity}

\subsubsection{Counterfactual Model Choice and Fresh Context}
\label{ssec:energy-tiers}

With token counts held fixed, Opus requires \num{10.2} times the energy of
Haiku, a ratio between \num{9.6} and \num{10.5} in every workstream although
their totals differ about \num{500}-fold (Table~\ref{tab:energy-tiers},
Appendix~\ref{app:energy}). Across vendors the range widens to a factor of
\num{62}, from Haiku~4.5 to Gemini~2.5~Pro (Table~\ref{tab:energy-vendors}),
but its upper end rests on a registry estimate of \numrange{200}{600}B active
parameters outside the calibrated range, and a fixed trace ignores that
vendors tokenize and cache the same text differently. These equal-trace
counterfactuals isolate the effect of model choice on a fixed workload; they
hold the turns fixed, whereas another model may need more or fewer turns, or
fail.

How sessions are used is a second lever. Had every episode
started a new session, dropping the context that each of the 691 episodes not
opening a session inherited from earlier ones, the workload would have read
\SI{76}{\percent} fewer cached tokens, and the hybrid estimate would fall from \num{248} to
\SI{97}{\kilo\watt\hour} (\SI{-61}{\percent}; \SIrange{-59}{-63}{\percent}
across the parameter bounds) and from \num{96} to
\SI{38}{\kilo\gram} CO$_2$eq (Appendix~\ref{app:fresh}); EcoLogits registers
no change. Because a fresh session loses information the task may need, and
recovering it costs calls the counterfactual does not add, the figure bounds
the saving from above.

\subsubsection{Which Factors Matter}
\label{ssec:energy-sensitivity}

Table~\ref{tab:factors} brings together every factor examined in this paper
and answers two questions. How much does a factor change the numbers? The
cached-context scenario, the attention term and the model tier each move the
total about tenfold, and the grid moves the carbon estimate 14-fold. Could it
change the conclusion? Only if it closed the gap between the estimators, and
none does: the ratio never falls to 1. Charging attention reads at the
reference energy per bit leaves \num{4.6}. Whether the attention term,
calibrated on short requests, holds at the contexts observed here cannot be
tested on the trace, but it can be bounded: even removing the attention reads
altogether leaves \num{1.4}, because prefill and attention computation over
the context remain. The
token-accounting findings are measured and depend on none of these factors.

Three of the largest uncertainties share a cause. The memory factors
$\eta(N)$ and $s_{\mathrm{attn}}(N)$ of the attention term were fitted under
a $d_{\mathrm{model}}$-sized key--value cache on dense 8B--72B models; we
reuse them with $d_{kv}$, and the upper parameter bounds (up to 200B) lie
outside the fitted range, where $\eta\,s_{\mathrm{attn}}$ rises from about
\num{15} at 72B to \num{60} at 200B. The term that carries \SI{90}{\percent}
of the estimate is therefore the least constrained by calibration, which is
why the attention reads, the key--value dimension and the parameter bound
rank high in group (a). The cache extension itself adds no free parameter
(Table~\ref{tab:app-extension}).

\begin{table}[t]
\centering
\caption{Every factor examined, varied one at a time from the baseline at the
low parameter bound (\SI{105.0}{\kilo\watt\hour}, \SI{41.1}{\kilo\gram}
CO$_2$eq, hybrid/EcoLogits ratio \num{14.5}, against \num{25} at the
central value). Each row changes one
assumption from the first setting to the second and gives, in the same order,
the hybrid total and the hybrid/EcoLogits ratio at both settings; the two
estimators would agree at a ratio of~1. Group~(a) changes the size of the
gap, group~(b) only the absolute values; within each group, factors are
ranked by effect. PUE 1.56: average reported in the Uptime Institute 2024
survey~\cite{uptime2024survey}; ---: not computed.}
\label{tab:factors}
\footnotesize
\setlength{\tabcolsep}{2.5pt}
\begin{tabular}{@{}llrr@{}}
\toprule
Factor & Setting & Hybrid total & Ratio \\
\midrule
\multicolumn{4}{@{}l}{\emph{(a) Changes the size of the gap}} \\
Cached context    & omitted $\to$ naive          & 18.5 $\to$ 232 kWh & 2.6 $\to$ 32 \\
Attention reads   & removed $\to$ modeled        & 10.1 $\to$ 105 kWh & 1.4 $\to$ 14.5 \\
Model tier        & Haiku $\to$ Opus             & 16.9 $\to$ 172 kWh & 73 $\to$ 14 \\
KV dimension      & $d_{kv}\to d_{\mathrm{model}}$ & 105 $\to$ 700 kWh & 14.5 $\to$ 97 \\
Active parameters & low $\to$ high               & 105 $\to$ 585 kWh  & 14.5 $\to$ 43 \\
Energy per bit    & H100 $\to$ model             & 33 $\to$ 105 kWh   & 4.6 $\to$ 14.5 \\
Fresh session     & per task $\to$ observed      & 43 $\to$ 105 kWh   & 5.9 $\to$ 14.5 \\
\midrule
\multicolumn{4}{@{}l}{\emph{(b) Changes only the absolute values}} \\
Grid (carbon)     & Sweden $\to$ Poland          & 4.4 $\to$ 62.5 kg  & 11 $\to$ 14 \\
PUE               & 1.2 $\to$ 1.56               & 105 $\to$ 136 kWh  & 14.5 $\to$ 19 \\
Batch size        & 64 $\to$ 4                   & 102 $\to$ 115 kWh  & 14 $\to$ 16 \\
Delegations       & accounted $\to$ reconstr.    & 105 $\to$ 110 kWh  & --- \\
Embodied (carbon) & excluded $\to$ included      & 40.4 $\to$ 41.1 kg & 14.5 $\to$ 14 \\
\midrule
\multicolumn{4}{@{}p{\columnwidth}@{}}{\emph{Not affected by any factor:} the
token accounting of Section~\ref{sec:results} (cache-read share, context
growth, delegated undercount), which is measured.} \\
\bottomrule
\end{tabular}
\end{table}

\subsubsection{Plausibility of the Attention Term}
\label{ssec:plausibility}

The decoding-time experiment supports the functional form of the attention
term: in three open-weight models, the time per output token grows almost
linearly with the context length ($R^2 = 0.98$ to $1.00$), and in the two
models whose context was served from cache, by the same amount as without it,
as Eq.~\eqref{eq:attn} assumes (Appendix~\ref{app:sweep}).
This supports the relation between context and decoding cost, not the
coefficient that converts it into energy. For the coefficient, a reference is the \SI{26.1}{\pico\joule\per\bit} that
an H100 spends when streaming memory at full power. The model charges
attention reads at \num{1.05} to \num{6.0} times this reference if
$s_{\mathrm{attn}}$ is an energy overhead, and at \num{0.6} to \num{1.3}
times it if $s_{\mathrm{attn}}$ is additional traffic
(Appendix~\ref{ssec:ceiling}); charging them at the reference instead lowers
the total by a factor of \num{3.2} under the first reading
(Table~\ref{tab:factors}) and by \SI{7}{\percent} under the second. The hybrid
figures therefore lie at or towards the high end of the plausible range.

%% file: sections/subagents.tex
\subsection{Accounting for Delegated Subagents}
\label{sec:subagents}

Delegation is infrequent: \num{44} of the \num{746} working episodes
(\SI{5.9}{\percent}) call a subagent. Appendix~\ref{app:subagents} describes
the three subagent types and what their runs do.

\subsubsection{Undercounting of Delegated Tokens}
\label{ssec:sub-trap}\label{ssec:sub-types}

The usage reported to the parent describes only the final internal API call
of a subagent (an example is given in Appendix~\ref{app:subagents}). A private
transcript was found for each of the \num{106} completed foreground
delegations. Over their \num{1334} internal calls (\num{12.6} per delegation),
the reported usage underestimates tokens by a factor of \num{10.0}
(\num{5970133} reported against \num{59771362}; by agent type in
Table~\ref{tab:subagents}). Per delegation, the median ratio is \num{7.4}, so
every delegation is undercounted, and the ratio is nearly a monotone function
of the number of internal calls (Spearman $\rho = 0.99$), as expected if only
the final call is reported. The missing tokens are mainly cache writes and
reads.

\subsubsection{Model Attribution and Energy}
\label{ssec:sub-model}\label{ssec:sub-result}

In \num{82} of the \num{106} delegations (\SI{77}{\percent}) the tool result
names no model, and the accounted workload assigns the parent's. The private
transcripts show that most delegated calls ran on Haiku~4.5, so inheriting the
parent's model overestimates energy per token while the final-call usage
underestimates tokens. Estimating each internal call separately with its
recorded model avoids both errors and yields \SI{7840}{\watt\hour} for the
\num{106} delegations, against \SI{4793}{\watt\hour} under reported usage: the
tenfold token undercount becomes a \num{1.6}-fold energy undercount, because
the reported final call carries the largest context of its run and most of the
work moves from Sonnet or Opus to Haiku. The \num{17} background subagents
report no usage; the transcripts of \num{8} of them (\num{104} internal calls,
\num{8299904} tokens, on Sonnet~5 and Opus~4.8) add \SI{2014}{\watt\hour}, and
\num{9} cannot be estimated. Together, the corrections raise the hybrid
estimate by only \SI{4.8}{\percent}, from \SI{105.0}{\kilo\watt\hour}
(accounted workload, low parameter bound) to \SI{110.0}{\kilo\watt\hour}
(reconstructed workload of \num{5161} requests), and
the delegated share of energy from \SI{4.6}{\percent} to \SI{9.0}{\percent};
the gap between the estimators, driven by main-thread cache reads, is
unaffected.

The undercount still matters for three reasons: the logging behavior is
undocumented and gives no sign of truncation; token-volume studies are fully
exposed to it, since delegated tokens are \SI{0.6}{\percent} of the trace as
logged but at least \SI{6.4}{\percent} in reality; and the partial
cancellation observed here depends on routing to Haiku and would not occur for
delegation to Sonnet or Opus.

%% file: sections/discussion.tex
\section{Discussion}\label{sec:discussion}

\subsection{Implications for Energy-Aware Agentic Software Engineering}

\paragraph*{Energy demand accumulates over the session} The
computational demands of agentic software engineering are shaped not only by
the model and the output it generates, but by how context is accumulated,
reused and propagated across tasks. Because an agent re-sends the
conversation with every tool call, cached-context reads make up
\SI{94.8}{\percent} of API-reported tokens, and context grows with the session
rather than with the task (Section~\ref{sec:results}). Each request therefore
carries the history of the whole session, and an estimate built on a single
request and response, as published per-query
footprints~\cite{elsworth2025google,oviedo2026joule} are, cannot describe
such a workload.

\paragraph*{Managing context can reduce energy} In the hybrid model,
attention reads over accumulated context carry most of the estimated energy,
and starting each task from a fresh context lowers the estimate by about
\SI{60}{\percent} while keeping the same requests and outputs
(Section~\ref{ssec:energy-tiers}). Because the estimate holds the requests
fixed, it bounds the saving from above; it places context isolation,
selective retention and compaction alongside conventional inference optimizations: the aim is to keep
the information a task needs without re-processing history it does not.

\paragraph*{Routing work to smaller models can reduce energy} On the same
trace, Opus needs about
ten times the energy of Haiku under the hybrid model, which favors routing
routine operations such as repository exploration to small models and
reserving larger ones for tasks that need them. The relevant objective, however,
is energy per successfully completed task rather than per call, since model
choice also affects whether, and after how many interactions, a task
succeeds.

\paragraph*{Token accounting must cover the whole workflow} Two parts of the
workload are
invisible to conventional token accounting: cached context, which output-only
estimators ignore, and delegated work, which parent-level logs undercount
tenfold (Section~\ref{sec:subagents}). Separating the four token classes per
request and reconstructing delegated runs from their own transcripts exposes
both. The difference this makes is structural: EcoLogits has no input that
responds to context, and no assumption we varied, including removing the
attention reads altogether, brings the two estimates together
(Section~\ref{ssec:energy-sensitivity}). Inference-energy assessments of
agentic workloads therefore need to represent cached-context processing and
the execution structure of autonomous agents.

\subsection{Limitations and Future Work}

The trace comes from one engineer using one commercial assistant, selected by
inspection (Appendix~\ref{app:data}), so its task mix and totals are
specific to this work. The two mechanisms behind the token findings do not
depend on the engineer: the client re-sends the conversation with every tool
call, and the provider serves the repeated prefix from a prompt
cache~\cite{gim2024promptcache,anthropic2026caching}. We therefore expect the
dominance of cache reads and the growth of context to hold for other users of
this client; whether they hold for other assistants remains to be tested, and
the subagent undercount depends on the client version. Token counts are billing records, deduplicated by two
independent extraction implementations that agree exactly
(Appendix~\ref{app:collapse}); energy and emissions are estimated, not
measured. The attention term extends a model calibrated on short, uncached
requests to proprietary architectures and far longer contexts. The
decoding-time experiment supports its dependence on context and the
sensitivity analysis bounds its effect, but neither establishes the absolute
energy of the hosted models. The
counterfactuals hold the recorded calls fixed and do not capture changes in
task performance. Future work should combine
workflow-level tracing with hardware-level energy measurement and with
controlled evaluations of context management and model routing that assess
energy alongside task completion.

%% file: sections/conclusion.tex
\section{Conclusion}\label{sec:conclusion}

A measured trace of \num{0.99} billion cumulative API-reported tokens from ten
weeks of agentic software engineering departs fundamentally from the
single-exchange model assumed by token-based energy estimators: about
\SI{95}{\percent} of tokens are cache reads, context grows 5.3-fold over a
session but only 1.4-fold within an episode, and parent-level logs undercount
delegated tokens tenfold. The cache-aware hybrid estimator yields about \SI{250}{\kilo\watt\hour} against about
\SI{10}{\kilo\watt\hour} for an output-only estimator, a 25-fold gap that no
assumption we varied closes.

These results translate into concrete steps. Developers can start a new
session, or clear or compact the context, between unrelated tasks and route
routine work to smaller models; tool vendors can log the complete usage of
delegated runs and give users per-request token or energy feedback; and
researchers can report all four token classes per request. Energy-aware
agentic software engineering begins with accounting for the context that
agents re-read.

%% file: sections/appendix.tex
\appendices

\section{Data Processing}
\label{app:data}

\subsection{Trace Selection}

The data were extracted retrospectively from the JSON-lines transcripts that
Claude Code writes during ordinary use; no instrumentation was introduced for
this study. The engineer used the assistant for design, integration, testing
and documentation and, in some of the products, implementation. The trace
was constructed from fifteen project directories assigned to the six
workstreams, of which thirteen were available at extraction; the other two
could not be located. We excluded a workstream unrelated to software
development, the directory associated with the analysis itself, and
directories judged to contain fewer than approximately one million tokens.
This threshold was applied by inspection, and neither its precise value nor
the excluded token volume was recorded; the absolute totals therefore describe
the trace as selected.

\subsection{API Call Reconstruction}
\label{app:collapse}\label{ssec:collapse}

Two deduplication steps are applied, in this order: copies from resumed
sessions are removed first, and the remaining records are then merged into
responses. Two independent extraction implementations, sharing no code, apply
both steps and agree exactly on the resulting counts (\num{3738} responses,
\num{994} prompts, \num{106} completed delegations).

\subsubsection{Duplicate Records Across Resumed Sessions}

When a session is resumed, the new transcript begins with a verbatim copy of
the transcript it continues, in which every record retains its identifier,
timestamp and usage and only the session identifier differs. Merging within a
session cannot detect these copies. Transcripts are therefore processed in
chronological order, and a record whose identifier has already occurred in an
earlier transcript is discarded. In this trace, three resumed sessions
copied \num{607} distinct API calls, one chain of them twice, for \num{752}
surplus responses; without their removal every token total would be
overstated by \SI{28}{\percent}. All counts in this paper, including
both columns of Table~\ref{tab:collapse}, are after this removal.

\subsubsection{Duplicate Records Within API Responses}

The transcript format writes one assistant record per content block of an API
response, so that thinking blocks, text blocks and individual tool calls each
receive a separate record with its own identifier and timestamp. Each of these
records contains an identical copy of the response's usage object. A response
comprising one thinking block and eight tool calls therefore appears as nine
records reporting identical token counts; the largest response in the trace
spans \num{17} records. Summing usage over records consequently inflates every
token total by the number of content blocks per response, and since each record
is individually well-formed, this inflation is not apparent from the data.

Records that share a session and a response identifier are therefore merged:
the first record of a response contributes its usage and defines the turn, and
subsequent records are folded into it. This reduces \num{7706} transcript
records to \num{3738} distinct responses, a factor of \num{2.06}; \num{15} of
these responses are locally generated synthetic records with zero usage, not
API calls, and are excluded, leaving \num{3723} API calls. Tool-call blocks on the folded records
represent distinct tool invocations, so tool and delegation counts are
accumulated over all records, whereas only the usage object is deduplicated.

\subsubsection{Verification of Reconstructed Usage}

Deduplicating tool calls together with usage objects would correct the token
totals but would introduce an error of opposite sign in the tool and delegation
counts. Table~\ref{tab:collapse} confirms that the two are handled separately:
tool calls and delegations are identical in both columns, whereas all token
totals are approximately halved.

\begin{table}[t]
\centering
\caption{Usage counted once per transcript record and once per distinct
response. Tool and delegation counts are read from content blocks, which do not
repeat, so they are identical in both columns---the check that the collapse
deduplicates usage and nothing else. The \num{3738} responses include \num{15}
synthetic records with zero usage.}
\label{tab:collapse}
\footnotesize
\begin{tabular}{lrr}
\toprule
 & Per transcript & Per distinct \\
 & record & response \\
\midrule
Units counted & \num{7706} & \num{3738} \\
\midrule
Uncached input   & \num{556594}     & \num{167541} \\
Output           & \num{9350343}    & \num{3618355} \\
Cache write      & \num{110574136}  & \num{48302053} \\
Cache read       & \num{1982069379} & \num{940461095} \\
Total tokens     & \num{2102550452} & \num{992549044} \\
\midrule
Cache-read share    & \SI{94.27}{\percent} & \SI{94.75}{\percent} \\
Cache read : output & 212 : 1              & 260 : 1 \\
\midrule
Tool calls          & \num{3710} & \num{3710} \\
Completed delegations & \num{106}  & \num{106} \\
\bottomrule
\end{tabular}
\end{table}

Token shares are largely insensitive to the correction---the cache-read share
moves from \SI{94.27}{\percent} to \SI{94.75}{\percent}---but ratios between
classes are not: each class inflates by a different factor (uncached input
\num{3.32}$\times$, output \num{2.58}$\times$, cache write \num{2.29}$\times$,
cache read \num{2.11}$\times$), so cache read to output moves from 212:1 to
260:1. The extraction script can regenerate the left-hand column of Table~\ref{tab:collapse} for any snapshot.

\subsection{Episode Identification and Classification}

An episode comprises one human request, a prompt or consecutive prompts
submitted before the assistant responds, and all assistant calls preceding the
next request; six of the 752 episodes are terminal prompts without subsequent
activity. Since the transcript's user role does not consistently identify
human authorship, human requests are identified by excluding tool-result and
automatic-notification messages: the explicit origin field of a message is
present for well under half of the \num{994} prompts, having been introduced
during the logging period. Without merging consecutive prompts,
\SI{26}{\percent} of episodes would contain a prompt but no assistant activity.

The declared task type is assigned by ordered keyword rules taken from an
existing internal session-logging taxonomy, applied first to the prompt and
then to the session title. The behavioral label is the episode's most frequent
tool category: reading, editing, command execution, web access, delegation
(subagent calls and skill invocations, which load predefined instruction
sets), other (task-list, planning and connector tools), or none for episodes
without tools. Ties (54 episodes) are resolved by the category encountered
first. The rule definitions are part of the analysis code (available on request), and the category
counts are given in Appendix~\ref{app:workload}. The \num{17} background
subagents are counted as delegations but excluded from per-subagent
statistics, since their tool result reports neither usage nor tool counts.

\section{Analytical Energy Model and Estimator Assumptions}
\label{app:equations}

Table~\ref{tab:app-notation} defines the quantities used in this appendix.
Constants are listed separately in Tables~\ref{tab:app-constants}
and~\ref{tab:app-eco}.

\begin{table}[t]
\centering
\caption{Notation. Token classes follow Section~\ref{sec:processing};
architectural quantities follow Section~\ref{sec:assumptions}.}
\label{tab:app-notation}
\footnotesize
\begin{tabular}{@{}ll@{}}
\toprule
Symbol & Meaning \\
\midrule
$T_{\mathrm{in}}$ & uncached input tokens \\
$T_{\mathrm{prompt}}$ & all input tokens (original model, no caching) \\
$T_{\mathrm{fresh}}$ & uncached input plus cache-write tokens \\
$T_{\mathrm{cached}}$ & cache-read tokens \\
$T_{\mathrm{ctx}}$ & context length, $T_{\mathrm{fresh}} + T_{\mathrm{cached}}$ \\
$T_{\mathrm{out}}$ & generated output tokens \\
$A(T)$ & key--value entries read while decoding after context $T$ \\
\midrule
$N$ & active parameter count \\
$N_{\mathrm{tot}}$ & total parameter count (sizes hardware only) \\
$N_{\mathrm{B}}$ & $N$ in billions \\
$n_{\mathrm{layers}}$ & transformer layers \\
$d_{\mathrm{model}}$ & hidden dimension \\
$d_{kv}$ & effective key--value dimension (KV heads $\times$ head dim.) \\
$b_{\mathrm{KV}}$ & bits per cached token, $2\,n_{\mathrm{layers}}\,d_{kv}\,b$ \\
\midrule
$B$ & serving batch size \\
$n_{\mathrm{GPU}}$ & accelerators holding the model \\
$\tau_{\mathrm{meas}}$, $\tau_{\mathrm{gen}}$ & measured latency, generation time \\
\bottomrule
\end{tabular}
\end{table}

\subsection{Original Model}
\label{app:original}

For a request with $T_{\mathrm{prompt}}$ input and $T_{\mathrm{out}}$ output
tokens, the analytical model of Vartziotis et al.~\cite{vartziotis2026tokens}
estimates the accelerator energy from the tensor-core operations of the
prefill and decode phases, $C_{\mathrm{pre}}$ and $C_{\mathrm{dec}}$, and the
traffic through high-bandwidth memory,
\begin{multline}
E_{\mathrm{GPU}}^{\mathrm{orig}} = \alpha_{\mathrm{TC}}\,(C_{\mathrm{pre}} + C_{\mathrm{dec}}) \\
+ e_{\mathrm{mem}}\,\eta(N)\,\bigl(\mathrm{Bits}_{\mathrm{param}}
+ \mathrm{Bits}_{\mathrm{KV}} + \mathrm{Bits}_{\mathrm{attn}}\bigr),
\label{eq:orig}
\end{multline}
where $\alpha_{\mathrm{TC}}$ is the energy per operation, $e_{\mathrm{mem}}$
the energy per bit of memory access, and $\eta(N)$ a memory-inefficiency
factor. The operation counts are
\begin{align}
C_{\mathrm{pre}} &= K N T_{\mathrm{prompt}}
  + 2\,n_{\mathrm{layers}}\,d_{\mathrm{model}}\,T_{\mathrm{prompt}}^{2}, \\
C_{\mathrm{dec}} &= K N T_{\mathrm{out}}
  + 4\,n_{\mathrm{layers}}\,d_{\mathrm{model}}\,A(T_{\mathrm{prompt}}),
\end{align}
with $A(T) = T_{\mathrm{out}}T + \tfrac{1}{2}T_{\mathrm{out}}(T_{\mathrm{out}}-1)$,
and the memory traffic is
\begin{align}
\mathrm{Bits}_{\mathrm{param}} &= b_w\,N\,\gamma(N), \\
\mathrm{Bits}_{\mathrm{KV}} &= 2\,b\,n_{\mathrm{layers}}\,d_{\mathrm{model}}\,T_{\mathrm{out}}, \\
\mathrm{Bits}_{\mathrm{attn}} &= s_{\mathrm{attn}}(N)\,
  2\,b\,n_{\mathrm{layers}}\,d_{\mathrm{model}}\,A(T_{\mathrm{prompt}}),
\end{align}
where $N$ is the parameter count of a dense model. The attention corrections of
$C_{\mathrm{pre}}$ and $C_{\mathrm{dec}}$ are included when the architecture is
known. The model was compared with measurements for dense models of 8B to 72B
parameters, with deviations of
\SIrange{5}{27}{\percent} for requests of 500 input and 500 output tokens~\cite{vartziotis2026tokens}. Its parameter-only
variant, which replaces the memory terms by an effective coefficient, is not
used here.

\subsection{Layer 1: Cache-Aware GPU Energy}

Layer~1 is the cache-aware GPU estimator of Section~\ref{sec:cache-model},
and Layer~2 (Appendix~\ref{app:layer2}) the infrastructure layer of
Section~\ref{sec:facility-carbon}. Layer~1 keeps the structure, constants and
calibrated factors of the original model; what is new is which token counts
and which key--value width enter each term, listed in
Table~\ref{tab:app-extension} and justified in Appendix~\ref{app:extension}.
The GPU energy of a request is the sum of a tensor-core compute term and a
memory-traffic term,
\begin{equation}
E_{\mathrm{GPU}}^{\mathrm{hyb}} = \alpha_{\mathrm{TC}}\,(C_{\mathrm{pre}} + C_{\mathrm{dec}})
+ e_{\mathrm{mem}}\,\eta(N)\,\mathrm{Bits}_{\mathrm{total}},
\label{eq:app-egpu}
\end{equation}
where $N$ is the number of active parameters. The compute terms comprise the
parameter-scaled transformer estimate with $K = 6$
operations per parameter and token~\cite{kaplan2020scaling,hoffmann2022training}
and the attention corrections of~\cite{vartziotis2026tokens},
\begin{align}
C_{\mathrm{pre}} &= K N T_{\mathrm{fresh}}
  + 2\,n_{\mathrm{layers}}\,d_{\mathrm{model}}\,T_{\mathrm{fresh}}\,T_{\mathrm{ctx}}, \\
C_{\mathrm{dec}} &= K N T_{\mathrm{out}}
  + 4\,n_{\mathrm{layers}}\,d_{\mathrm{model}}\,A(T_{\mathrm{ctx}}).
\end{align}
Only $T_{\mathrm{fresh}}$ enters the dense prefill term, whereas the attention
corrections run over the full context. The memory traffic consists of three
parts,
\begin{align}
\mathrm{Bits}_{\mathrm{total}} &= \mathrm{Bits}_{\mathrm{param}} + \mathrm{Bits}_{\mathrm{KV}}
  + \mathrm{Bits}_{\mathrm{attn}}, \\
\mathrm{Bits}_{\mathrm{param}} &= b_w\,N\,\gamma(N), \\
\mathrm{Bits}_{\mathrm{KV}} &= b_{\mathrm{KV}}\,(T_{\mathrm{fresh}} + T_{\mathrm{out}}), \\
\mathrm{Bits}_{\mathrm{attn}} &= s_{\mathrm{attn}}(N)\,b_{\mathrm{KV}}\,A(T_{\mathrm{ctx}})\label{eq:bitsattn},
\end{align}
with $b_{\mathrm{KV}}$ as in Eq.~\eqref{eq:attn}. In the notation of
Eq.~\eqref{eq:hybrid}, $E_{\mathrm{prefill}} = \alpha_{\mathrm{TC}}C_{\mathrm{pre}}$
and $E_{\mathrm{decode}} = \alpha_{\mathrm{TC}}C_{\mathrm{dec}}$, each including
its attention-compute correction, and $E_{\mathrm{param}}$, $E_{\mathrm{KV}}$
and $E_{\mathrm{attn}}$ are the three memory parts multiplied by
$e_{\mathrm{mem}}\,\eta(N)$. The parameter-access term
follows the original formulation~\cite{vartziotis2026tokens} and is charged
once per request, independently of output length. It represents an effective
memory-energy contribution rather than an explicit reconstruction of parameter
transfers during autoregressive decoding. In practice, model weights may be
accessed repeatedly across decoding steps, with the associated traffic
amortized through batching and weight reuse. These effects are not explicitly
represented by the term, and the calibration of the original model as a whole
leaves this allocation unconstrained for the proprietary serving
configurations considered here. Under the alternative convention the attention reads
still dominate: reading the full weights at every
decoding step without reuse, shared across the batch, would raise the
parameter term from \SI{4.8}{\watt\hour} to \SI{1.7}{\kilo\watt\hour} at the
baseline batch size of \num{16} and to \SI{26.7}{\kilo\watt\hour} at batch
size \num{1}, and lower the attention share of GPU energy from
\SI{93}{\percent} to \SI{92}{\percent} and \SI{71}{\percent}, respectively,
since the weights are shared across a batch whereas each request reads its own
context. Cache reads contribute no
memory term of their own; their entries are read in
$\mathrm{Bits}_{\mathrm{attn}}$, whose context length $T_{\mathrm{ctx}}$
includes them.

\begin{table}[t]
\centering
\caption{Constants of Layer~1~\cite{vartziotis2026tokens}. The energy
coefficients are the total (control plus datapath) values measured on the
Hopper GPU of an NVIDIA GH200 (Table~3 of~\cite{antepara2025benchmark}):
HBM energy per bit for $e_{\mathrm{mem}}$ and FP16 matrix-unit energy per
FLOP for $\alpha_{\mathrm{TC}}$; the memory
factors were fitted to the measurements of~\cite{caravaca2025prompts}.}
\label{tab:app-constants}
\footnotesize
\begin{tabular}{lll}
\toprule
Symbol & Value & Meaning \\
\midrule
$K$                    & 6                     & operations per parameter and token \\
$\alpha_{\mathrm{TC}}$ & \SI{0.52}{\pico\joule} per op. & tensor-core energy \\
$e_{\mathrm{mem}}$     & \SI{11.68}{\pico\joule\per\bit} & HBM energy per bit \\
$b_w$, $b$             & \num{16} bits         & weight and KV element width \\
\midrule
$\gamma(N)$            & $\gamma_0 (N/N_0)^{\beta}$     & parameter-access factor \\
$\gamma_0$, $\beta$, $N_0$ & \num{0.10}, \num{0.8}, 24B & \\
$s_{\mathrm{attn}}(N)$ & $1 + 1.5\,(N/N_0)^{0.9}$       & attention-read overhead \\
$\eta(N)$              & $1 + 0.8\,(N/N_0)^{0.8}$       & memory inefficiency \\
\bottomrule
\end{tabular}
\end{table}

\subsection{Derivation of the Extension}
\label{app:extension}

\begin{table*}[t]
\centering
\caption{Terms of the original model~\cite{vartziotis2026tokens} and of
Layer~1, with $b_{\mathrm{KV}}^{\mathrm{orig}} = 2\,n_{\mathrm{layers}}\,d_{\mathrm{model}}\,b$
and $b_{\mathrm{KV}} = 2\,n_{\mathrm{layers}}\,d_{kv}\,b$.
For $T_{\mathrm{cached}} = 0$, $d_{kv} = d_{\mathrm{model}}$ and without the
key--value writes of prefill, both columns coincide.}
\label{tab:app-extension}
\footnotesize
\begin{tabular}{@{}lll@{}}
\toprule
Term & Original model & Layer~1 \\
\midrule
Dense prefill compute     & $K N T_{\mathrm{prompt}}$
                          & $K N T_{\mathrm{fresh}}$ \\
Prefill attention compute & $2\,n_{\mathrm{layers}}\,d_{\mathrm{model}}\,T_{\mathrm{prompt}}^{2}$
                          & $2\,n_{\mathrm{layers}}\,d_{\mathrm{model}}\,T_{\mathrm{fresh}}\,T_{\mathrm{ctx}}$ \\
Dense decode compute      & $K N T_{\mathrm{out}}$
                          & $K N T_{\mathrm{out}}$ \\
Decode attention compute  & $4\,n_{\mathrm{layers}}\,d_{\mathrm{model}}\,A(T_{\mathrm{prompt}})$
                          & $4\,n_{\mathrm{layers}}\,d_{\mathrm{model}}\,A(T_{\mathrm{ctx}})$ \\
Parameter traffic         & $b_w N \gamma(N)$, $N$ all parameters
                          & $b_w N \gamma(N)$, $N$ active parameters \\
Key--value writes         & $b_{\mathrm{KV}}^{\mathrm{orig}}\,T_{\mathrm{out}}$
                          & $b_{\mathrm{KV}}\,(T_{\mathrm{fresh}} + T_{\mathrm{out}})$ \\
Attention reads           & $s_{\mathrm{attn}}(N)\,b_{\mathrm{KV}}^{\mathrm{orig}}\,A(T_{\mathrm{prompt}})$
                          & $s_{\mathrm{attn}}(N)\,b_{\mathrm{KV}}\,A(T_{\mathrm{ctx}})$ \\
\bottomrule
\end{tabular}
\end{table*}

Table~\ref{tab:app-extension} compares each term of the original model with
the corresponding term of Layer~1. Dense decode computation and parameter
traffic are unchanged; all other terms differ only in the token counts or the
key--value width that enter them. Three changes matter for the estimate; two
further ones are negligible.

\paragraph{Fresh versus cached computation}
In the original model, every input token attends over the full input, giving
$T_{\mathrm{prompt}}^{2}$, and every input token enters the dense prefill term.
Under caching, only the $T_{\mathrm{fresh}}$ processed tokens are computed, and
each of them attends over the full context $T_{\mathrm{ctx}}$, giving
$T_{\mathrm{fresh}}T_{\mathrm{ctx}}$. Decode attention compute and attention
reads run over $T_{\mathrm{ctx}}$, since cached entries are read at every
decoding step. The convention of the original model, which counts the full
context for every prefill token, is retained.

\paragraph{Key--value dimension and calibration transfer}
The equations of the original model express key--value traffic with
$d_{\mathrm{model}}$, whereas its text refers to an effective key--value
dimension $d_{kv}$. For grouped-query attention, $d_{kv}$ is the number of
key--value heads times the head dimension, e.g.\ $8 \times 128 = 1024$ for
Llama-3-70B against $d_{\mathrm{model}} = 8192$. The use of $d_{kv}$ is an
architectural correction: it reflects the size of the entries actually stored.
The memory factors $\eta(N)$ and $s_{\mathrm{attn}}(N)$, however, are retained
from the original calibration, which employed $d_{\mathrm{model}}$. The
extension is therefore not independently calibrated. Because the attention
reads carry \SI{90}{\percent} of the estimate (Table~\ref{tab:energy-terms}),
this is its least constrained assumption; the $d_{\mathrm{model}}$ reading
raises the trace total by a factor of \num{6.7}
(Table~\ref{tab:factors}). The decoding-time experiment of
Appendix~\ref{app:sweep} supports $d_{kv}$ as the width of the traffic but
cannot test whether the factors transfer; this transfer is therefore the
largest source of uncertainty in the absolute estimates, whereas the
token-accounting results do not depend on it.

\paragraph{Active parameters}
The original model treats dense models and names mixture-of-experts
architectures as future work~\cite{vartziotis2026tokens}. In a
mixture-of-experts model, each token is processed by the routed subset of
parameters, so the operation count and the per-token memory reads scale with
the active count. Layer~1 therefore uses the active count throughout, and the
total count enters only the number of GPUs in Layer~2. If all experts were
loaded from memory for every request, the parameter traffic would have to be
evaluated with the total count; this would raise the trace total by about
\SI{1.5}{\percent}.

\paragraph{Minor terms}
The original model charges key--value writes to the generated tokens only.
Prefill, however, writes the entries of every token it processes, which is the
precondition for caching them, so the extension charges
$T_{\mathrm{fresh}} + T_{\mathrm{out}}$; the term contributes less than
\SI{0.01}{\percent} of the energy of the present trace. Retrieving cached
entries into accelerator memory is not modeled as a separate term; charged as
one full memory read of the cached entries per request, it would add about
\SI{0.02}{\percent} to the trace total.

\subsection{Proof of the Scenario Ordering}
\label{app:bounds}

We prove Eq.~\eqref{eq:bounds}. The inequalities order the three scenarios
within the model; they are not measured bounds on hardware energy. Write
$E = E_{\mathrm{GPU}}^{\mathrm{hyb}}(T_{\mathrm{fresh}}, T_{\mathrm{cached}}, T_{\mathrm{out}})$
with $T_{\mathrm{ctx}} = T_{\mathrm{fresh}} + T_{\mathrm{cached}}$. For the
lower bound, fix $T_{\mathrm{fresh}}$ and $T_{\mathrm{out}}$. Then
$T_{\mathrm{cached}}$ enters only through $T_{\mathrm{ctx}}$ in the prefill and
decode attention compute and in the attention reads, each of which is linear
in $T_{\mathrm{ctx}}$ with a non-negative coefficient. Hence $E$ is
non-decreasing in $T_{\mathrm{cached}}$, and setting $T_{\mathrm{cached}} = 0$
gives the lower bound. For the upper bound, fix $T_{\mathrm{ctx}}$ and
$T_{\mathrm{out}}$ and reassign the cached tokens to the fresh class. Dense
prefill compute, prefill attention compute and key--value writes are
non-decreasing in $T_{\mathrm{fresh}}$ at fixed $T_{\mathrm{ctx}}$, and all
other terms depend only on $T_{\mathrm{ctx}}$ and $T_{\mathrm{out}}$. Hence
$E \le E_{\mathrm{GPU}}^{\mathrm{hyb}}(T_{\mathrm{ctx}}, 0, T_{\mathrm{out}})$.
Layer~2 depends on output tokens and latency only, so both inequalities also
hold for facility energy and, at fixed grid intensity, for emissions.

\subsection{Layer 2: Infrastructure}\label{app:layer2}

Layer~2 follows the EcoLogits methodology~\cite{ecologits2025} with its
constants (Table~\ref{tab:app-eco}). The number of GPUs is determined by the
memory required to hold all weights, so the total parameter count
$N_{\mathrm{tot}}$ enters here, whereas all other terms use the active count,
\begin{equation}
n_{\mathrm{GPU}} = 2^{\lceil \log_2 \lceil M / M_{\mathrm{GPU}} \rceil \rceil},
\qquad M = 1.2 \cdot N_{\mathrm{tot}} \cdot b_w / 8,
\end{equation}
with $M$ in bytes and $M_{\mathrm{GPU}} = \SI{80}{\giga\byte}$. The
generation time is the minimum of the measured latency and a per-token latency
regression,
\begin{equation}
\tau_{\mathrm{gen}} = \min\bigl(T_{\mathrm{out}}\,(\alpha_L N_{\mathrm{B}}
+ \beta_L B + \gamma_L),\; \tau_{\mathrm{meas}}\bigr).
\end{equation}
The non-GPU server energy and the embodied emissions are allocated
according to the generation time, the share of a server occupied by the
model and the batch size,
\begin{align}
E_{\mathrm{server}} &= \frac{\tau_{\mathrm{gen}}}{B}\,P_{\mathrm{server}}\,
  \frac{n_{\mathrm{GPU}}}{n_{\mathrm{srv}}}, \\
\mathrm{GWP}_{\mathrm{emb}} &= \frac{\tau_{\mathrm{gen}}}{B\,t_{\mathrm{life}}}
  \Bigl(\frac{n_{\mathrm{GPU}}}{n_{\mathrm{srv}}}\,G_{\mathrm{srv}}
  + n_{\mathrm{GPU}}\,G_{\mathrm{GPU}}\Bigr).
\end{align}

The GPU energy of the EcoLogits baseline, Eq.~\eqref{eq:eco}, is a regression
on decode-throughput benchmarks applied per GPU,
\begin{equation}
E_{\mathrm{GPU}}^{\mathrm{Eco}} = T_{\mathrm{out}}
  \bigl(\alpha_E\,e^{\beta_E B} N_{\mathrm{B}} + \gamma_E\bigr)\,n_{\mathrm{GPU}}
  \quad [\si{\watt\hour}].
\end{equation}
Since $\beta_E < 0$, the energy per output token decreases with
the batch size. It replaces Layer~1 in the EcoLogits estimate and is the only term in which
the two estimators differ structurally; apart from it, they differ only in
batch size and PUE, whose effects are quantified in Appendix~\ref{app:estimator}.

\begin{table}[t]
\centering
\caption{Constants of Layer~2 and of the EcoLogits GPU term, from the EcoLogits
v0.11.1 source~\cite{ecologits2025}, together with the defaults of this study
(Section~\ref{sec:assumptions}).}
\label{tab:app-eco}
\footnotesize
\begin{tabular}{lll}
\toprule
Symbol & Value & Meaning \\
\midrule
$\alpha_L$ & $6.785\times10^{-4}$ \si{\second} & latency per token and B param. \\
$\beta_L$  & $3.119\times10^{-4}$ \si{\second} & latency per token and batch slot \\
$\gamma_L$ & $1.947\times10^{-2}$ \si{\second} & latency per token, offset \\
$P_{\mathrm{server}}$ & \SI{1.2}{\kilo\watt} & non-GPU server power \\
$n_{\mathrm{srv}}$ & \num{8} & GPUs per server \\
$t_{\mathrm{life}}$ & \num{3} years & hardware lifetime \\
$G_{\mathrm{srv}}$ & \SI{5700}{\kilo\gram} CO$_2$eq & server, excluding GPUs \\
$G_{\mathrm{GPU}}$ & \SI{273}{\kilo\gram} CO$_2$eq & per GPU \\
\midrule
$\alpha_E$ & $1.167\times10^{-6}$ & GPU energy, slope \\
$\beta_E$  & $-1.121\times10^{-2}$ & GPU energy, batch exponent \\
$\gamma_E$ & $4.053\times10^{-5}$ & GPU energy, offset \\
\midrule
$B$ & \num{16} & batch size (EcoLogits: \num{64}) \\
PUE & \num{1.2} & power usage effectiveness \\
$I_{\mathrm{grid}}$ & \num{0.3844} & \si{\kilo\gram} CO$_2$eq/\si{\kilo\watt\hour}, US \\
\bottomrule
\end{tabular}
\end{table}

\subsection{Estimator Assumptions}
\label{app:estimator}

\paragraph{Model architecture and identifier mapping}
Layer count and hidden width follow from the active-parameter estimate with
the aspect ratio of Llama-3-70B and
$N \approx 12\,n_{\mathrm{layers}}\,d_{\mathrm{model}}^{2}$. The effective KV
dimension $d_{kv}=1024$ (eight KV heads of dimension 128) follows the
grouped-query attention of the Llama-3 models from 8B to 405B parameters. For
mixture-of-experts models, active parameters enter the computational and
per-request memory terms and total parameters the hardware-capacity estimate.
The compute and memory coefficients remain the GH200 (Hopper) values of the
original model, with 16-bit KV entries. Sonnet~5 and Opus~5, absent from the
registry, are mapped to Sonnet~4.6 and Opus~4.8; the substitutions are applied
identically in both estimators and do not contribute to the difference between
them. None of these proxies establishes the actual serving configuration of
the proprietary systems.

\paragraph{Latency, attribution and infrastructure}
The latency proxy is the time since the preceding transcript record for
main-thread calls (median \SI{5.446}{\second}) and the reported duration for
foreground delegations; it may include waiting and tool activity, and stalled
delegations are capped at \SI{1800}{\second} (seven calls, the longest
\SI{75579}{\second}). In both estimators, the generation time that allocates
server and embodied impacts is the minimum of this proxy and the EcoLogits
regression prediction. The analytical GPU term represents the model as a whole
and is not multiplied by the number of serving GPUs; the estimated GPU count
only allocates server activity and embodied impacts. If a parent-reported
delegation lacks a model identifier (82 of 106), the accounted workload
attributes it to the nearest preceding main-thread model in the same session;
the reconstructed workload uses each internal call's recorded model.

The parameter-scenario range excludes uncertainty from serving hardware,
attention architecture, quantization, calibration transfer and model-specific
optimizations. The hybrid default $\mathrm{PUE} = 1.2$ is close to the provider PUE of \numrange{1.09}{1.14} that EcoLogits
assumes for Anthropic and below the survey average of \num{1.56} reported by
the Uptime Institute~\cite{uptime2024survey}; the latter is examined as a
sensitivity setting (Table~\ref{tab:factors}). EcoLogits retains $B = 64$.

Batch size and PUE cannot be shared between the estimators because EcoLogits
does not expose them. Since batch size enters the hybrid model only through
its infrastructure layer (Layer~2, Appendix~\ref{app:layer2}), re-estimating it at the EcoLogits value of \num{64} instead of \num{16}
changes its total by \SI{2.4}{\percent} (from \num{105.0} to
\SI{102.4}{\kilo\watt\hour} at the lower parameter bound). Rescaling EcoLogits
from its provider PUE (\numrange{1.09}{1.14}) to $\mathrm{PUE} = 1.2$ increases
its total from \num{9.9} to \SI{10.6}{\kilo\watt\hour} and reduces the central
energy ratio from \num{25.1} to \num{23.3}. Neither difference accounts for a
substantial part of the gap between the estimators.

\section{Plausibility of the Attention Term}
\label{app:sweep}

\subsection{Decoding-Time Experiment}

\begin{figure*}[t]
  \centering
  \includegraphics[width=\textwidth]{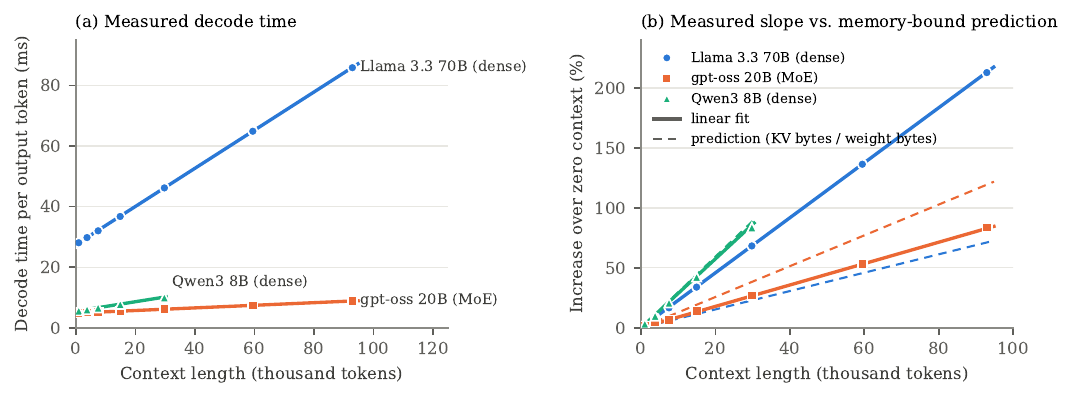}
  \caption{Decode time per output token against context length for three
  open-weight models with at most 256 output tokens. (a) Medians over three
  repetitions, with linear fits; the medians of uncached and repeated runs
  coincide within \SI{2}{\percent}. (b) Increase over the fitted zero-context time; solid: measured fit,
  dashed: prediction from the ratio of key--value bytes per context token to
  weight bytes per decoding step, taken from each model's configuration (for
  gpt-oss-20b only the full-attention layers grow with the context).}
  \label{fig:sweep}
\end{figure*}

Since energy cannot be metered on the vendor's service, we tested the
mechanism behind $E_\mathrm{attn}$ (Eq.~\eqref{eq:attn}) through decoding time, which is
limited by memory bandwidth and should therefore grow linearly with the
context. Three open-weight models, Llama~3.3~70B, gpt-oss-20b and Qwen3-8B
(Ollama tags \texttt{llama3.3:latest}, \texttt{gpt-oss:20b} and
\texttt{qwen3:8b}, with their default 4-bit weights and a 16-bit key--value
cache), were served by Ollama on a shared server with two NVIDIA H100 NVL
GPUs, at batch size 1. The context was varied from about \num{500} to
\num{93000} tokens for Llama, from \num{1000} to \num{93000} for gpt-oss-20b
and from \num{1000} to \num{30000} for Qwen3-8B, whose context window is
about \num{40000} tokens, with three repetitions and at most 256 output tokens
(gpt-oss-20b stopped after 85 to 155). Each prompt consisted of newly
generated random text and was sent twice: the first run had to process it in
full, and the second, identical run could reuse the cached prefix. Server-side
timers separate prefill from decode time.
In all three models the time per output token grows linearly
with the context ($R^2 = 0.98$ to $1.00$), with a slope within a factor of
three of the memory-bound prediction (Fig.~\ref{fig:sweep}). At about
\num{93000} tokens, reading the context takes \SI{68}{\percent} (Llama) and
\SI{45}{\percent} (gpt-oss-20b) of each decoding step, against
\SIrange{1}{3}{\percent} at about \num{1000} tokens. For gpt-oss-20b and
Qwen3-8B, the repeated run took \SIrange{0.2}{9}{\percent} of the original
prefill time, whereas its median decode time per token stayed within
\SI{2}{\percent} of the uncached run (individual runs \SI{-6}{\percent} to
\SI{+7}{\percent}): caching removes the prefill time but not this per-token
cost, as assumed by Layer~1. Llama~3.3~70B was reloaded before 46 of its 48
requests on the shared server, which discards the cache, so almost none of
its repeated runs were cached and it contributes to the linearity result
only.

The predicted slope depends on the key--value width used to size the traffic.
With each model's actual key--value width ($d_\mathrm{kv}$, the number of
key--value heads times the head dimension), the measured slope is 0.69 to 3.0
times the prediction; with the hidden dimension $d_\mathrm{model}$, as in the
equations of~\cite{vartziotis2026tokens}, the prediction exceeds the measurement
by a factor of 2.7 to 8.3 for all three models (Table~\ref{tab:kvwidth}). The
context-dependent traffic therefore scales with $d_\mathrm{kv}$, as assumed by
Layer~1. For Qwen3-8B, the memory bandwidth implied by the zero-context step
time and by the context slope agree (\num{910} and \SI{939}{\giga\byte\per\second}),
as expected if both are limited by memory reads. For Llama~3.3~70B they differ
by a factor of three (\num{1558} and \SI{520}{\giga\byte\per\second}), the
deviation also seen in Table~\ref{tab:kvwidth}; given the repeated reloads
under memory pressure, its placement on the GPUs may have varied between
requests, and the deviation is not interpreted further. For gpt-oss-20b, whose
key--value width is 512 and
whose sliding-window layers do not grow with the context, the fixed
$d_\mathrm{kv}=1024$ of Layer~1 would overstate the slope by a factor of about
three. The experiment does not test whether $\eta(N)$ and $s_\mathrm{attn}(N)$,
fitted under $d_\mathrm{model}$, transfer to $d_\mathrm{kv}$.

The check therefore supports the computational behavior that
Eq.~\eqref{eq:attn} encodes, namely that every decoding step reads the full
context, cached or not, at a cost linear in its length. It does not validate
the energy coefficient $e_{\mathrm{mem}}\,\eta(N)\,s_{\mathrm{attn}}(N)$ that
converts this traffic into energy, whose calibration transfer from
$d_{\mathrm{model}}$ to $d_{kv}$ is discussed in Appendix~\ref{app:extension};
Appendix~\ref{ssec:ceiling} gives a reference value for its scale. The
experiment further measures time rather than energy, at batch size 1 and on
a different serving stack than the calibration, and its longest context lies
below the median context of the trace's requests (\num{153138} tokens).

\begin{table}[t]
  \centering
  \caption{Measured context slope of decode time divided by the slope predicted
  from the key--value bytes read per context token (1 = exact), for the two
  readings of the key--value width, settings as in the text; for gpt-oss-20b
  only the 12 full-attention layers grow with the context.}
  \label{tab:kvwidth}
  \begin{tabular}{lrrrr}
    \toprule
    Model & $d_\mathrm{kv}$ & $d_\mathrm{model}$ & \multicolumn{2}{c}{Measured / predicted} \\
          &                 &                    & with $d_\mathrm{kv}$ & with $d_\mathrm{model}$ \\
    \midrule
    Llama 3.3 70B & 1024 & 8192 & 3.0  & 0.37 \\
    gpt-oss-20b   &  512 & 2880 & 0.69 & 0.12 \\
    Qwen3-8B      & 1024 & 4096 & 0.97 & 0.24 \\
    \bottomrule
  \end{tabular}
\end{table}

\subsection{Reference Energy per Bit}
\label{ssec:ceiling}

The decoding-time experiment tests the functional form of the attention term,
not its coefficient. For the coefficient, a reference value indicates the
scale. An H100 SXM drawing its full board power of \SI{700}{\watt} while
streaming memory at its peak bandwidth of \SI{3.35}{\tera\byte\per\second}
spends \SI{26.1}{\pico\joule\per\bit}. This is not a physical upper bound,
because the energy per bit also depends on the bandwidth actually achieved,
but it marks what a fully loaded accelerator spends on streaming memory. The
hybrid model charges $e_{\mathrm{mem}}\,\eta(N)$ per bit of key--value traffic
inflated by $s_{\mathrm{attn}}(N)$. If $s_{\mathrm{attn}}$ is an energy
overhead on a single read per decoding step, the charge per stored bit at the
lower bounds is \num{27}, \num{96} and \SI{157}{\pico\joule} for Haiku, Sonnet
and Opus, \num{1.05}, \num{3.7} and \num{6.0} times the reference; if
$s_{\mathrm{attn}}$ represents real additional traffic, it is \num{16},
\num{27} and \SI{33}{\pico\joule}. As a sensitivity setting, charging the
attention reads at the reference value instead lowers the trace total from
\num{105.0} to \SI{33.3}{\kilo\watt\hour} under the first reading and to
\SI{98.1}{\kilo\watt\hour} under the second (Table~\ref{tab:factors} lists
the first).
Since the model charges at or above what a fully loaded accelerator spends
under the first reading, and close to it under the second, the hybrid figures
are more likely to lie at or towards the high end of the plausible range, and
absolute estimates should be reported together with the key--value
formulation and the parameter bounds.

\section{Supplementary Workload Characterization}
\label{app:workload}

\subsection{Workload Composition and Task Characteristics}

\begin{figure*}[t]
\centering
\includegraphics[width=\textwidth]{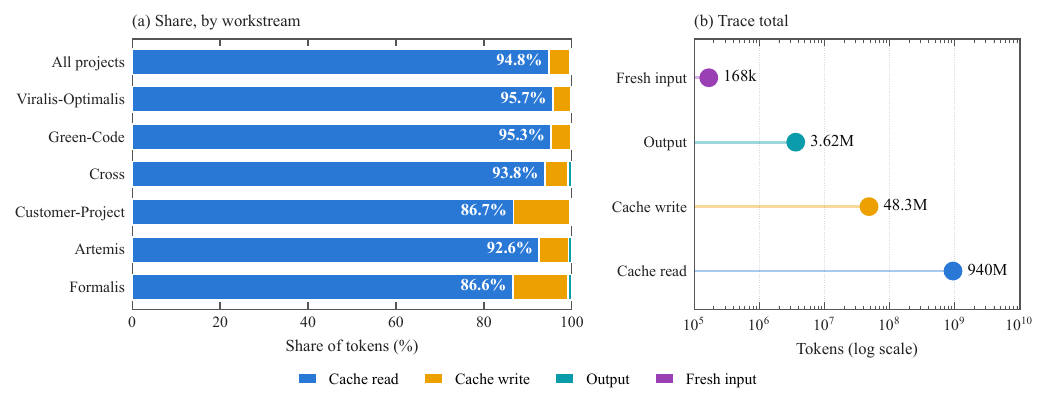}
\caption{Token composition: (a) share of each token class by workstream;
(b) trace-wide absolute totals per class, log scale. Cache reads dominate every
workstream.}
\label{fig:composition}
\end{figure*}

Fig.~\ref{fig:composition} shows the token composition of each workstream,
and Table~\ref{tab:tasks} lists the episodes, median tokens, median output
and median turns of each declared task type.

\begin{table}[t]
\centering
\caption{The \num{746} working episodes by declared type. Medians are per episode;
$k$ denotes thousands of tokens.}
\label{tab:tasks}
\footnotesize
\begin{tabular}{lrrrrr}
\toprule
Task type & $n$ & \% & Tok.\,$k$ & Out.\,$k$ & Turns \\
\midrule
Feature     & 270 & 36.2 & \num{314} & 1.5 & 2 \\
Understand  & 154 & 20.6 & \num{182} & 0.9 & 1 \\
Integration &  82 & 11.0 & \num{490} & 2.2 & 1 \\
Docs        &  53 &  7.1 & \num{439} & 2.0 & 3 \\
Bugfix      &  42 &  5.6 & \num{269} & 1.5 & 3 \\
Other       &  39 &  5.2 & \num{166} & 0.9 & 1 \\
Review      &  34 &  4.6 & \num{311} & 2.1 & 3 \\
Test        &  26 &  3.5 & \num{298} & 2.9 & 4.5 \\
Init        &  24 &  3.2 & \num{317} & 2.9 & 3 \\
Design      &  15 &  2.0 & \num{501} & 3.1 & 2 \\
Refactor    &   7 &  0.9 & \num{835} & 2.4 & 1 \\
\bottomrule
\end{tabular}
\end{table}

\paragraph{Task label versus episode length and position}
To test whether the task-type differences reflect the label or other
properties of the episode, we regress log total tokens of the \num{732} working
episodes with non-zero usage on (i) the declared task type alone, (ii) log turn
count, log position of the episode in its session and the primary model, and
(iii) both. The coefficients of determination are \num{0.063}, \num{0.889} and
\num{0.893}. The task label thus adds less than half a percentage point once
episode length, session position and model are known; most of the variation
it explains on its own is shared with these variables. The regression is
descriptive: turn count is itself an outcome of the task, and episodes within a
session are not independent.

\subsection{Task Labels and Observed Tool Behavior}

Of the \num{746} working episodes, the behavioral label is none for \num{342}
(\SI{45.8}{\percent}), command execution for \num{128}, reading for \num{107},
editing for \num{105}, other for \num{32}, delegation for \num{24} and web
access for \num{8}; among the \num{154} Understand episodes, \num{112}
(\SI{72.7}{\percent}) invoke no tool.

The two axes agree in part. Bugfix episodes mostly run commands
(\SI{31}{\percent}), Docs mostly edit files (\SI{28}{\percent}), and Test
mostly reads (\SI{38}{\percent}). Review is the only type whose most common
signature is delegation (\SI{32}{\percent}), though six of those eleven
episodes invoked a skill rather than a subagent. Counting subagent calls alone,
review is still the task most often handed off rather than done inline: \num{8}
of the \num{34} Review episodes (\SI{24}{\percent}) called a subagent, the
highest rate of any type.

\subsection{Interaction Structure and Model Routing}

The trace holds \num{994} prompts and \num{3738} assistant turns, so one
prompt produces \num{3.76} turns and \num{1.00}M tokens on average. The
median number of turns per episode varies by task type (Table~\ref{tab:tasks}),
from \num{4.5} for Test to one for Understand, Integration, Other and Refactor.

\begin{figure}[t]
\centering
\includegraphics[width=\columnwidth]{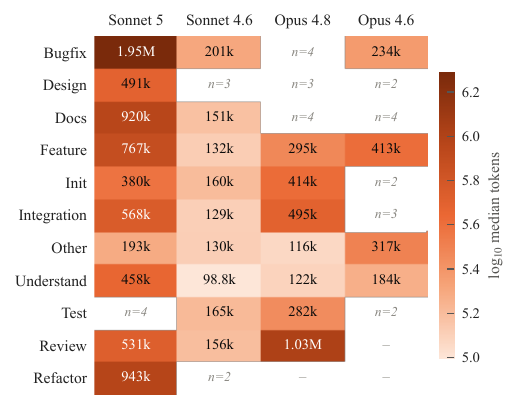}
\caption{Median tokens per episode by task type and primary model, as routed during the trace.}
\label{fig:model}
\end{figure}

Fig.~\ref{fig:model} crosses task type with the model that ran most of the
episode. We did not control model selection, so the figure describes what
happened rather than how the models compare. A model used mainly for hard
problems will show a high median for that reason alone. Cells with fewer than
five episodes show their count instead of a median, and we drop three models
with fewer than \num{20} episodes.

\section{Supplementary Energy, Carbon, and Sensitivity Results}
\label{app:energy}

Two reporting conventions are used in this appendix. The per-workstream
comparison of the estimators (Table~\ref{tab:energy-family},
Fig.~\ref{fig:e3})
reports central values, the geometric mean of the two active-parameter bounds,
for a hybrid trace total of \SI{247.7}{\kilo\watt\hour}. All decompositions
and model counterfactuals
(Tables~\ref{tab:energy-terms}--\ref{tab:energy-tiers},
Figs.~\ref{fig:e7} and~\ref{fig:e12}) use the lower bound, for a total of
\SI{105.0}{\kilo\watt\hour} (Section~\ref{sec:assumptions}). Values from the
two groups are therefore not directly comparable.

\subsection{Estimator Comparison Across Workstreams}
\label{ssec:energy-where}

\begin{figure*}[t]
\centering
\includegraphics[width=\textwidth]{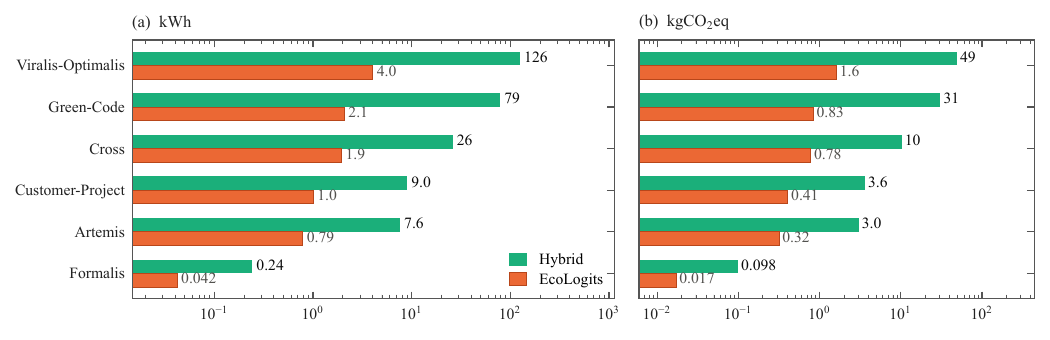}
\caption{Energy (a) and carbon (b) per workstream under both estimators, log
scale, central values. Workstreams are ordered by hybrid energy. On the log axis
the horizontal gap between each pair of bars is the hybrid-to-EcoLogits ratio of
Table~\ref{tab:energy-family}.}
\label{fig:e3}
\end{figure*}

The difference between the estimators is present in all six workstreams but
varies in magnitude, from a factor of \num{5.8} in \wsPC{} to \num{38.0} in
\wsRA{} (Table~\ref{tab:energy-family}, Fig.~\ref{fig:e3}). It is associated
with cache-read intensity rather than with call volume: the customer project has
nearly as many calls as \wsRB{} (\num{607} against \num{662}) but a ratio of
\num{9.0} against \num{13.5}, and \wsPB{} (\num{427} calls) one of \num{9.7},
because their sessions accumulate shorter contexts.

The token composition shows the underlying coverage problem. Of the
\num{0.999} billion tokens of the accounted workload, comprising the \num{0.99} billion main-thread
tokens (Table~\ref{tab:composition}) and \num{5970133} delegated tokens,
\SI{94.71}{\percent} are cache reads and \SI{4.87}{\percent} cache writes.
Output tokens, the only class captured by Eq.~\eqref{eq:eco}, account for
\SI{0.405}{\percent}, and uncached input for \SI{0.017}{\percent}.

\begin{table*}[t]
\centering
\caption{Per workstream, central values. The ratio column is
hybrid energy divided by EcoLogits energy. Calls and tokens include the
\num{106} delegated subagent calls, with their reported usage, and exclude the
\num{15} synthetic records, so they exceed the turn and token counts of
Table~\ref{tab:corpus}; for \wsRA{}, for example, the \num{916} turns less
\num{3} synthetic records plus \num{15} delegations give \num{928} calls.}
\label{tab:energy-family}
\footnotesize
\begin{tabular}{lrrrrr}
\toprule
Workstream & Calls & Tokens & \multicolumn{2}{c}{Energy (\si{\watt\hour})} & Ratio \\
\cmidrule(lr){4-5}
 & & & EcoLogits & Hybrid & \\
\midrule
\wsPA{} & \num{1147} & \num{477411480} & \num{4034.8} & \num{125697.3} & 31.2$\times$ \\
\wsRA{}        & \num{928}  & \num{333632151} & \num{2078.4} & \num{78976.4}  & 38.0$\times$ \\
\wsRB{}        & \num{662}  & \num{85889425}  & \num{1941.7} & \num{26138.0}  & 13.5$\times$ \\
\wsCu{}          & \num{607}  & \num{57208752}  & \num{1003.6} & \num{9032.5}   & 9.0$\times$ \\
\wsPB{}         & \num{427}  & \num{41801671}  & \num{787.5}  & \num{7637.3}   & 9.7$\times$ \\
\wsPC{}         & \num{58}   & \num{2575698}   & \num{41.9}   & \num{242.6}    & 5.8$\times$ \\
\bottomrule
\end{tabular}
\end{table*}

\subsection{Energy Decomposition}

Table~\ref{tab:energy-terms} lists the hybrid estimate by physical term; the
sensitivity of the total to each modeling assumption is given in
Table~\ref{tab:factors}. In the cache-naive scenario, the \num{0.95} billion
cache-read tokens of the accounted workload undergo a full forward pass, which raises GPU energy from
\num{84.6} to \SI{190.7}{\kilo\watt\hour} (factor \num{2.25}).

\begin{table}[t]
\centering
\caption{Hybrid energy by term, low active-parameter bound. Facility energy is
IT energy $\times$ PUE.}
\label{tab:energy-terms}
\footnotesize
\begin{tabular}{lrr}
\toprule
Term & \si{\watt\hour} & Share \\
\midrule
Memory: attention reads   & \num{79040.6}  & \SI{90.36}{\percent} \\
Compute: prefill (fresh)  & \num{5050.6}   & \SI{5.77}{\percent} \\
Non-GPU server            & \num{2855.9}   & \SI{3.26}{\percent} \\
Compute: decode           & \num{523.8}    & \SI{0.60}{\percent} \\
Memory: parameter loads   & \num{4.8}      & \SI{0.01}{\percent} \\
Memory: KV writes         & \num{1.0}      & $<\SI{0.01}{\percent}$ \\
\midrule
IT total                  & \num{87476.7}  & \SI{100}{\percent} \\
Facility ($\times 1.2$)   & \num{104972.0} & \\
\bottomrule
\end{tabular}
\end{table}

\subsection{Dependence on Input, Context and Output Tokens}
\label{ssec:energy-vs-output}

\begin{figure*}[t]
\centering
\includegraphics[width=\textwidth]{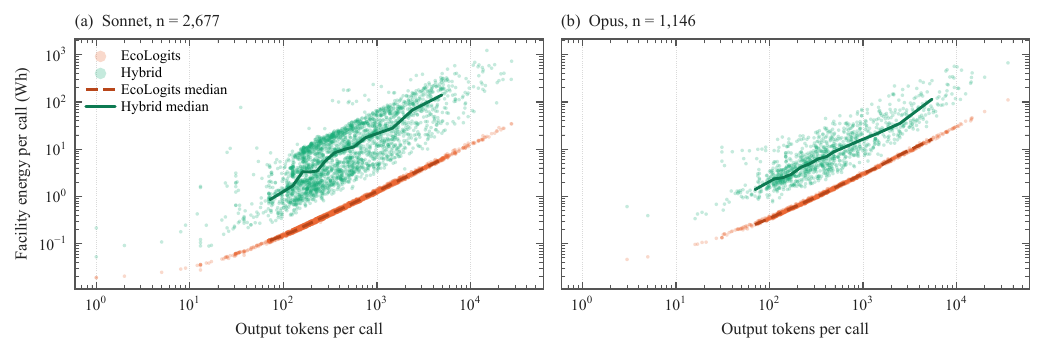}
\caption{Per-call facility energy against output tokens, both estimators at the
low active-parameter bound, faceted by model tier, log--log. Lines are binned
medians (dashed: EcoLogits). At fixed output count (5th--95th percentile, median
over ten output bins) the EcoLogits points span a factor of \num{1.4}, whereas
the hybrid estimate spans \num{14.3} (Sonnet) and \num{8.4} (Opus). Haiku is
omitted (\num{6} calls).}
\label{fig:e8}
\end{figure*}

\begin{figure*}[t]
\centering
\includegraphics[width=\textwidth]{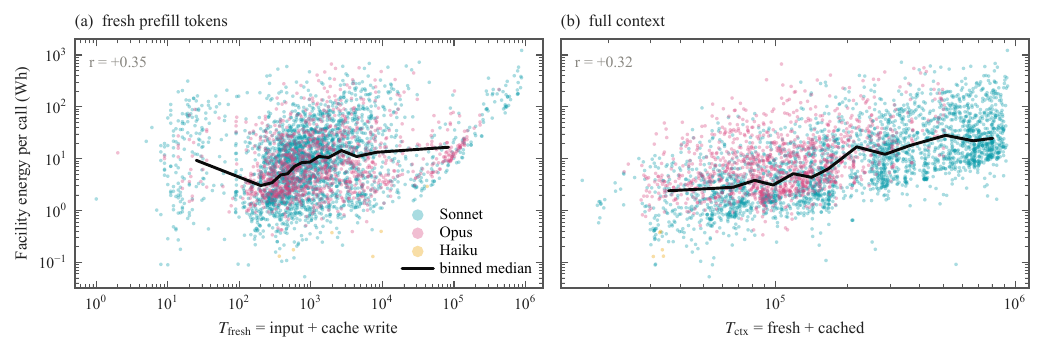}
\caption{Per-call hybrid facility energy at the low active-parameter bound
against fresh prefill tokens (a) and full context (b), log--log, colored by
model tier. The line is a binned median, not a fit. Raw
uncached input is omitted: median \num{2}, Spearman $\rho = 0.34$.}
\label{fig:e7}
\end{figure*}

Fig.~\ref{fig:e8} plots per-call energy against output tokens, and
Fig.~\ref{fig:e7} against $T_{\mathrm{fresh}}$ and $T_{\mathrm{ctx}}$,{} as
binned medians without a parametric fit. Over the \num{3829} calls, the
Spearman rank correlations of hybrid energy with uncached input,
$T_{\mathrm{fresh}}$, $T_{\mathrm{ctx}}$ and output are \num{0.34},
\num{0.28}, \num{0.59} and \num{0.81} (Pearson on linear values: \num{0.03},
\num{0.35}, \num{0.32} and \num{0.78}, dominated by the largest calls). The
spread within each bin spans orders of magnitude, so no single token
quantity characterizes the energy of a call.

\subsection{Counterfactual Model Comparisons}
\label{ssec:energy-models}

\paragraph{Claude tiers}
Table~\ref{tab:energy-tiers} gives the energy of each workstream on Haiku,
Sonnet and Opus. Actual usage sits between the all-Sonnet and all-Opus
cases, which is what a Sonnet-dominant mix should give.

\paragraph{Three vendors}
Table~\ref{tab:energy-vendors} adds GPT-5.4 and Gemini~2.5~Pro, with the
caveats of Section~\ref{ssec:energy-tiers}: registry parameter estimates,
the same token counts despite different tokenizers, and, for Gemini, a
parameter range outside the calibration.

\begin{table}[t]
\centering
\caption{The accounted workload re-estimated on current models from three vendors,
tokens held fixed. Energy is facility energy at the low active-parameter
bound. The Gemini row lies outside the calibrated parameter range.}
\label{tab:energy-vendors}
\footnotesize
\begin{tabular}{llrr}
\toprule
Model & Vendor & Active (B) & \si{\kilo\watt\hour} \\
\midrule
Haiku 4.5      & Anthropic & \numrange{10}{35}   &   16.9 \\
Sonnet 4.6     & Anthropic & \numrange{44}{132}  &   92.5 \\
GPT-5.4        & OpenAI    & \numrange{45}{150}  &   96.0 \\
Opus 4.8       & Anthropic & \numrange{67}{200}  &  172.1 \\
Gemini 2.5 Pro & Google    & \numrange{200}{600} & 1042.2 \\
\midrule
\multicolumn{3}{l}{as actually run (mixed)}      &  105.0 \\
\bottomrule
\end{tabular}
\end{table}

\begin{table}[t]
\centering
\caption{The same tokens re-estimated on each Claude tier, facility
\si{\kilo\watt\hour} at the low active-parameter bound. Registry
active-parameter ranges: Haiku \numrange{10}{35}B, Sonnet \numrange{44}{132}B,
Opus \numrange{67}{200}B.}
\label{tab:energy-tiers}
\footnotesize
\setlength{\tabcolsep}{3pt}
\begin{tabular}{lrrrrr}
\toprule
Workstream & Haiku & Sonnet & Opus & actual & O/H \\
\midrule
\wsPA{} & 8.37 & 45.42 & 84.46 & 52.96 & 10.1$\times$ \\
\wsRA{}        & 6.10 & 33.33 & 62.07 & 33.65 & 10.2$\times$ \\
\wsRB{}        & 1.53 &  8.61 & 16.12 & 10.99 & 10.5$\times$ \\
\wsCu{}          & 0.43 &  2.30 &  4.19 &  3.96 &  9.6$\times$ \\
\wsPB{}         & 0.48 &  2.70 &  5.04 &  3.30 & 10.5$\times$ \\
\wsPC{}         & 0.02 &  0.11 &  0.21 &  0.11 & 10.3$\times$ \\
\midrule
Trace             & 16.9 & 92.5 & 172.1 & 105.0 & 10.2$\times$ \\
\bottomrule
\end{tabular}
\end{table}

\begin{figure*}[t]
\centering
\includegraphics[width=\textwidth]{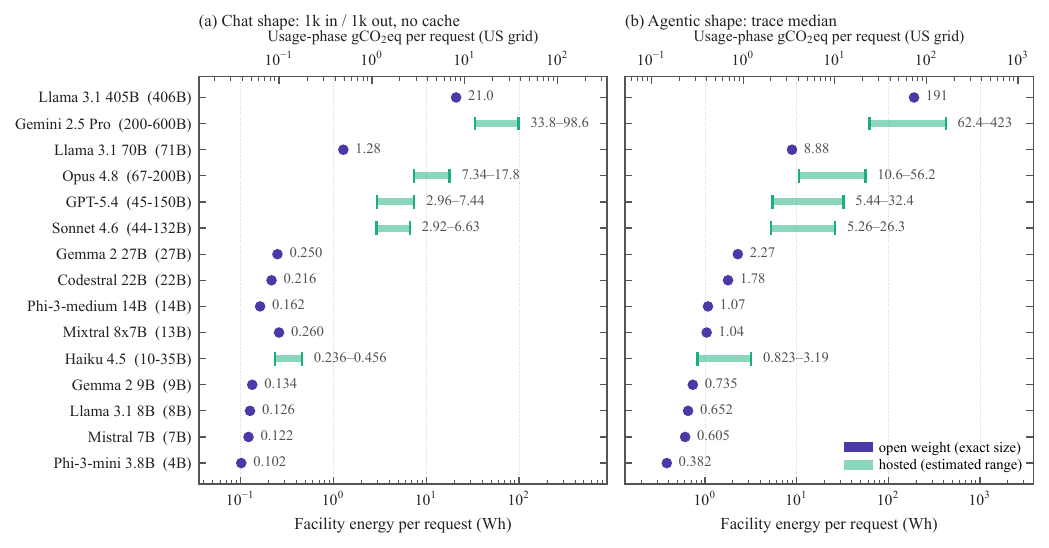}
\caption{Facility energy for one identical request across fifteen models, under
a chat-shaped workload (a) and this trace's median agentic workload (b), log
scale, with usage-phase carbon (embodied excluded) on the upper axis. Open-weight
models are points because their parameter counts are published; hosted models
are bars spanning the registry's estimated range, labeled low--high. Models are
ordered by active parameter count.}
\label{fig:e12}
\end{figure*}

\paragraph{One request, fifteen models}
Fig.~\ref{fig:e12} holds a single request fixed and varies only the model, for
two request shapes: a chat shape (\num{1000} uncached input and \num{1000} output
tokens, no caching) and an agentic shape built from the per-class medians of
the \num{3829} calls (\num{2} uncached input tokens, \num{838} cache writes,
\num{143098} cache reads, \num{387} output tokens). At the lower
active-parameter bound, estimated energy varies across models by a factor of
\num{333} for the chat shape and \num{500} for the agentic shape, the larger
range arising because the attention term scales with both context length and
model size. The comparison shows how the request shape interacts with model
size in the analytical model; it is not an efficiency ranking. Hosted models
rest on third-party parameter estimates spanning up to a factor of \num{3.5}
(bars), all models are evaluated with the same GH200 (Hopper) coefficients, and a smaller
model that needs more attempts to complete a task does not save
proportionally.

\subsection{Fresh-Context Counterfactual}
\label{app:fresh}

The counterfactual of Section~\ref{ssec:energy-tiers} rewrites the token
counts of every main-thread call and re-estimates the accounted workload with
everything else unchanged. For each session $s$, with resumed sessions merged,
the base context $c_s$ is the context $T_{\mathrm{ctx}}$ of its first call:
system prompt, tool definitions, project instructions and the first request,
which any new session would also carry (median \num{31300} tokens, range
\numrange{27400}{55800}). For every later episode $e$ of $s$, the inherited
context is
\begin{equation}
k_e = \max\bigl(0,\; T_{\mathrm{ctx}}(\text{first call of } e) - c_s\bigr),
\end{equation}
which slightly overstates the inheritance by the new prompt of $e$; its median
is \num{88400} tokens, and inherited context makes up \SI{83}{\percent} of the
summed context at episode starts. Each call of $e$ loses $k_e$ from its
context, but not below $\min(T_{\mathrm{ctx}}, c_s)$; fresh input and output
are unchanged, so the removed tokens come out of the cache reads. On the first
call of $e$, the base context is moved from cache reads to cache writes,
because a new session has to process it again. The 106 delegations are left
unchanged. Of the 732 episodes with priced calls, 691 do not open a session,
and all of them have $k_e > 0$. Over the accounted workload, cache reads fall
from
\num{945.7} to \num{228.2} million tokens and cache writes rise from \num{48.6}
to \num{67.4} million. The facility energy falls from \num{105.0} to
\SI{43.0}{\kilo\watt\hour} at the low bound, from \num{584.7} to
\SI{217.5}{\kilo\watt\hour} at the high bound and from \num{247.7} to
\SI{96.8}{\kilo\watt\hour} at the central values. The script
\texttt{fresh\_context\_counterfactual.py} reproduces these numbers.

\subsection{Carbon and Deployment Scenarios}

\paragraph{Embodied emissions}
Usage-phase emissions exceed embodied emissions under both estimators, although
to different degrees: at the central values, embodied emissions account for
\SI{4.03}{\percent} of the EcoLogits total and \SI{0.76}{\percent} of the
hybrid total (\SI{1.7}{\percent} at the lower bound,
Table~\ref{tab:factors}), since the same
latency-amortized hardware share is compared with a larger usage term in the
latter case.{} In the fifteen-model comparison, embodied emissions add
\SIrange{1}{14}{\percent} for the agentic shape and \SIrange{29}{52}{\percent}
for the chat shape, for which the latency-amortized hardware share is
relatively larger.
\paragraph{Grid intensity}
For a fixed grid factor, usage-phase carbon is proportional to energy (factor
\num{0.3844}) and is therefore shown as a secondary axis in Fig.~\ref{fig:e12}.
For a fixed workload, the choice of model changes emissions by a factor of
\num{500} at the lower parameter bound and up to \num{1107} at the upper bound,
whereas the national grids of Section~\ref{ssec:energy-total} differ by a factor
of \num{17} (\SI{35}{\gram}/\si{\kilo\watt\hour} in Sweden to
\SI{589}{\gram}/\si{\kilo\watt\hour} in Poland).
Both figures are conditional on the registry parameter estimates and on the
assumed grid factor.

\section{Reconstruction and Accounting of Delegated Subagents}
\label{app:subagents}

\subsection{Parent-Reported Versus Internal Token Usage}

The usage a delegation reports to its parent is that of its final internal
call, not the sum over its run (Section~\ref{ssec:sub-trap}). In one
delegation of the trace, the eight internal calls read \num{0}, \num{24643},
\num{27303}, \num{37061}, \num{42882}, \num{46530}, \num{51688} and
\num{54079} cached tokens. The tool result reports \num{54079}, the value of
the final call, whereas the run read \num{284186} cached tokens in total, a
factor of \num{5.3} more.

\begin{table}[t]
\centering
\caption{The \num{106} foreground delegations by agent type. Int.: internal
API calls in the private transcripts; tokens (millions) as reported in the
parent's tool result and as summed over the internal calls; Sec.: median
duration; Edits: file edits.}
\label{tab:subagents}
\footnotesize
\setlength{\tabcolsep}{3pt}
\begin{tabular}{lrrrrrrr}
\toprule
Agent type & Calls & Int. & Rep. & Actual & Ratio & Sec. & Edits \\
\midrule
Explore         & 57 &  \num{891} & 3.28 & 37.84 & 11.5$\times$ & 107 &  0 \\
general-purpose & 38 &  \num{260} & 1.64 & 11.08 &  6.8$\times$ &  56 & 41 \\
Plan            & 11 &  \num{183} & 1.05 & 10.85 & 10.3$\times$ & 577 &  0 \\
\midrule
Total           & 106 & \num{1334} & 5.97 & 59.77 & 10.0$\times$ & & 41 \\
\bottomrule
\end{tabular}
\end{table}

Across the \num{106} foreground delegations, the undercount ranges from
\num{6.8} for general-purpose to \num{11.5} for Explore
(Table~\ref{tab:subagents}); per delegation, the ratio has an interquartile
range of \numrange{4.2}{10.8} and a range of \numrange{1.4}{35.8}.

\subsection{Delegated Work}

\begin{figure*}[t]
\centering
\includegraphics[width=\textwidth]{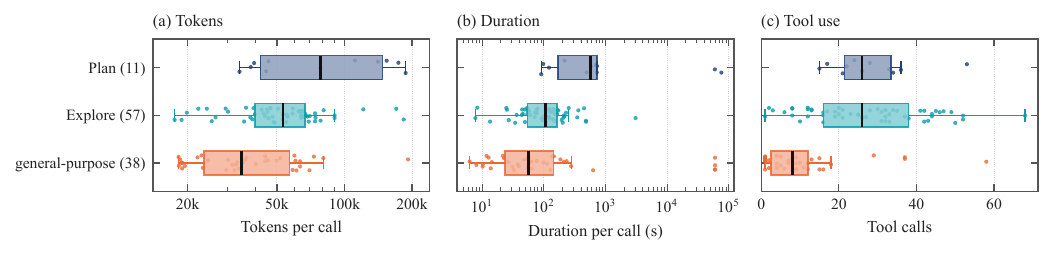}
\caption{Distribution of tokens (a), duration (b) and tool calls (c) per
subagent call, by subagent type; (a) and (b) on logarithmic axes. Boxes span the
interquartile range, the black line marks the median and whiskers extend to
1.5 times the interquartile range; points are individual calls, and the number
of calls is given in parentheses. Six calls (two Plan, four general-purpose)
lasted longer than one hour. File edits are given in
Table~\ref{tab:subagents}.}
\label{fig:subagents}
\end{figure*}

\begin{figure}[t]
\centering
\includegraphics[width=\columnwidth]{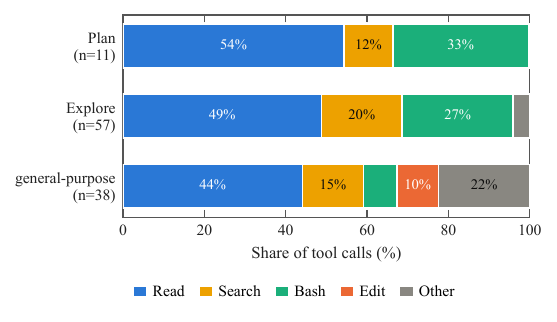}
\caption{Share of tool calls by category and subagent type (n = calls). Other: planning, task-list and connector tools. Unlike the episode-level
categories of Section~\ref{sec:processing}, search is counted separately from
reading. Only the general-purpose agent edits files.}
\label{fig:toolmix}
\end{figure}

The \num{106} completed foreground delegations fall in \num{36} of the
\num{44} working episodes that call a subagent; the other \num{8} contain only delegations that returned no
usage (Section~\ref{ssec:sub-model}). Claude Code provides three subagent
types: Explore for read-only codebase search, Plan for designing
implementation plans, and general-purpose with access to all tools. Their
recorded tool use matches these roles{} (Figs.~\ref{fig:subagents}
and~\ref{fig:toolmix}): Explore devotes \SI{49}{\percent} of its tool calls to
reading and \SI{20}{\percent} to searching and edits no file, general-purpose
performs all \num{41} file edits (\num{282} lines added, \num{202} removed),
and Plan has the highest median cost (\num{78}k tokens) and duration
(\SI{577}{\second}).

